%% file: main.tex
\documentclass[letterpaper,twocolumn,10pt]{article}
\usepackage{xcolor}
\usepackage{usenix2019_v3}

\usepackage{tikz}
\usepackage{amsmath}
\usepackage{amsthm}
\usepackage{xspace}
\usepackage{booktabs}
\usepackage{graphicx}
\usepackage{float}
\usepackage{subfigure}
\usepackage{pgfplots}
\usepackage{pgfplotstable}
\usepackage{threeparttable}
\usepackage{multicol}
\usepackage{lipsum,adjustbox}
\usepackage{multirow}
\usepackage{pgf-pie}
\usepackage{pgfkeys}
\usepackage{framed}
\usepackage{listings}
\usepackage{makecell}
\usepackage{appendix}
\usepackage{tabularx}
\usepackage{cleveref}
\usepackage[available,functional,reproduced]{usenixbadges}

\crefname{section}{}{}
\creflabelformat{section}{#2\S#1#3}
\creflabelformat{subsection}{#2\S#1#3}
\creflabelformat{subsubsection}{#2\S#1#3}

\input{listing-rustfmt}

\begin{document}

\date{}

\input{macro}

\title{\Large \bf \aster: A Linux ABI-Compatible, Rust-Based Framekernel OS\\ with a Small and Sound TCB}

\author{
{\rm
Yuke Peng\textsuperscript{1,}\blackthanks{Co-first authors.}\ ,
Hongliang Tian\textsuperscript{2,}\footnotemark[1]\ ,
Junyang Zhang\textsuperscript{3,4},
Ruihan Li\textsuperscript{3,4},
Chengjun Chen\textsuperscript{2},
}
\\
{\rm
Jianfeng Jiang\textsuperscript{2},
Jinyi Xian\textsuperscript{1},
Xiaolin Wang\textsuperscript{3,4},
Chenren Xu\textsuperscript{3,4},
Diyu Zhou\textsuperscript{3,4},
}
\\
{\rm
Yingwei Luo\textsuperscript{3,4,}\blackthanks{Co-corresponding authors: Yinqian Zhang (yinqianz@acm.org); Shoumeng Yan (shoumeng.ysm@antgroup.com); Yingwei Luo (lyw@pku.edu.cn).}\ ,
Shoumeng Yan\textsuperscript{2,}\footnotemark[2]\ ,
Yinqian Zhang\textsuperscript{1,}\footnotemark[2]\ 
}
\\
\textsuperscript{1}SUSTech \hspace{1cm}
\textsuperscript{2}Ant Group \hspace{1cm}
\textsuperscript{3}Peking University \hspace{1cm}
\textsuperscript{4}Zhongguancun Laboratory
} 

\maketitle

\begin{abstract}
\input{0-abstract}
\end{abstract}

\input{1-intro}

\input{2-background}

\input{3-framekernel}

\input{4-ostd}

\input{5-aster}

\input{6-eval}

\input{7-related}

\input{8-conclusion}

\input{9-acknowledgement}

\bibliographystyle{plain}
\bibliography{main}

\input{appendix}

\end{document}

%% file: listing-rustfmt.tex
\definecolor{commentgreen}{RGB}{0, 128, 0}
\definecolor{keywordblue}{RGB}{0, 0, 255}
\definecolor{stringpurple}{RGB}{128, 0, 128}

\lstdefinelanguage{Rust}{
  keywords={break, const, continue, crate, else, enum, extern, false, fn, for, if, impl,
    in, let, loop, match, mod, move, mut, pub, ref, return, self, Self, static, struct,
    super, trait, true, type, unsafe, use, where, while, async, await, dyn, abstract,
    become, box, do, final, macro, override, priv, try, typeof, unsized, virtual,
    yield, union},
  keywordstyle=\color{keywordblue}\bfseries,
  ndkeywords={bool, char, i8, i16, i32, i64, i128, isize, u8, u16, u32, u64, u128, usize,
    f32, f64, String, Vec, Result, Option, Box, Copy, Clone, Debug, Eq, Ord, PartialEq,
    PartialOrd, ToString, From, Into, Default, Read, Write},
  ndkeywordstyle=\color{keywordblue}\bfseries,
  identifierstyle=\color{black},
  sensitive=true,
  comment=[l]{//},
  morecomment=[s]{/*}{*/},
  commentstyle=\color{commentgreen}\ttfamily,
  stringstyle=\color{stringpurple}\ttfamily,
  morestring=[b]{"},
  morestring=[b]{'},
}

%% file: macro.tex
\newcommand{\bheading}[1]{{\vspace{3pt}\noindent{\textbf{#1}}}}

\newcounter{note}[section]
\renewcommand{\thenote}{\thesection.\arabic{note}}
\newcommand{\yz}[1]{\refstepcounter{note}{\bf\textcolor{red}{$\ll$YZ~\thenote: {\sf #1}$\gg$}}}
\newcommand{\jyx}[1]{\refstepcounter{note}{\bf\textcolor{red}{$\ll$JX~\thenote: {\sf #1}$\gg$}}}
\newcommand{\pyk}[1]{\refstepcounter{note}{\bf\textcolor{red}{$\ll$YK~\thenote: {\sf #1}$\gg$}}}
\newcommand{\zjy}[1]{\refstepcounter{note}{\bf\textcolor{orange}{$\ll$ZJY~\thenote: {\sf #1}$\gg$}}}

\newcommand{\major}[1]{#1}
\newcommand{\minor}[1]{#1}

\newcommand{\secref}[1]{\mbox{Sec.~\ref{#1}}\xspace}
\newcommand{\secrefs}[2]{\mbox{Sec.~\ref{#1}--\ref{#2}}\xspace}
\newcommand{\figref}[1]{\mbox{Fig.~\ref{#1}}}
\newcommand{\tabref}[1]{\mbox{Table~\ref{#1}}}
\newcommand{\appref}[1]{\mbox{Appendix~\ref{#1}}}
\newcommand{\ignore}[1]{}
\newcommand{\ssecref}[1]{\mbox{\S\ref{#1}}\xspace}
\newcommand{\etc}{\textit{etc.}\xspace}
\newcommand{\ie}{\textit{i.e.}\xspace}
\newcommand{\eg}{\textit{e.g.}\xspace}
\newcommand{\cf}{\textit{cf.}\xspace}
\newcommand{\aka}{\textit{a.k.a.}\xspace}
\newcommand{\etal}{\textit{et al.}\xspace}

\newcommand{\framekernel}{\textsc{FrameKernel}\xspace}
\newcommand{\framework}{\textit{Framework}\xspace}
\newcommand{\services}{\textit{Services}\xspace}
\newcommand{\asterinas}{\textsc{Asterinas}\xspace}

\newcommand{\ostd}{\textsc{Ostd}\xspace}
\newcommand{\aster}{\textsc{Asterinas}\xspace}
\newcommand{\kernmiri}{\textsc{KernMiri}\xspace}

\newcommand{\upperroman}[1]{\MakeUppercase{\romannumeral#1}}

\newcommand{\unsafekeyword}{\texttt{unsafe}\xspace}
\newcommand{\hpriv}{privileged\xspace}
\newcommand{\lpriv}{de-privileged\xspace}

\newcommand{\gbytes}{\ensuremath{\mathrm{GB}}\xspace}
\newcommand{\mbytes}{\ensuremath{\mathrm{MB}}\xspace}
\newcommand{\kbytes}{\ensuremath{\mathrm{KB}}\xspace}
\newcommand{\bytes}{\ensuremath{\mathrm{B}}\xspace}
\newcommand{\hertz}{\ensuremath{\mathrm{Hz}}\xspace}
\newcommand{\ghertz}{\ensuremath{\mathrm{GHz}}\xspace}
\newcommand{\msecs}{\ensuremath{\mathrm{ms}}\xspace}
\newcommand{\usecs}{\ensuremath{\mathrm{\mu{}s}}\xspace}
\newcommand{\nsecs}{\ensuremath{\mathrm{ns}}\xspace}
\newcommand{\secs}{\ensuremath{\mathrm{s}}\xspace}
\newcommand{\gbits}{\ensuremath{\mathrm{Gb}}\xspace}
\newcommand{\throuput}{\ensuremath{\mathrm{TPS}}\xspace}

\newcommand{\cmark}{\ding{51}}%
\newcommand{\xmark}{\ding{55}}%

\newcounter{packednmbr}
\newenvironment{packedenumerate}{
\begin{list}{\thepackednmbr.}{\usecounter{packednmbr}
\setlength{\itemsep}{0pt}
\addtolength{\labelwidth}{4pt}
\setlength{\leftmargin}{12pt}
\setlength{\listparindent}{\parindent}
\setlength{\parsep}{3pt}
\setlength{\topsep}{3pt}}}{\end{list}}

\newenvironment{packeditemize}{
\begin{list}{$\bullet$}{
\setlength{\labelwidth}{0pt}
\setlength{\itemsep}{2pt}
\setlength{\leftmargin}{\labelwidth}
\addtolength{\leftmargin}{\labelsep}
\setlength{\parindent}{0pt}
\setlength{\listparindent}{\parindent}
\setlength{\parsep}{1pt}
\setlength{\topsep}{1pt}}}{\end{list}}

\newtheorem{theorem}{Theorem}
\newtheorem{lemma}{Lemma}
\newtheorem{corollary}{Corollary}
\newtheorem{proposition}{Proposition}
\newtheorem{example}{Example}
\newtheorem{remark}{Invariant}
\newtheorem{definition}{Definition}
\newcommand{\e}[1]{{\mathbb E}\left[ #1 \right]}

\newtheoremstyle{invstyle}
{3pt}
{3pt}
{\itshape}
{}
{\bfseries}
{:}
{.5em}
{}

\theoremstyle{invstyle}
\newtheorem{inv}{Inv.}

\newcommand{\tabincell}[2]{\begin{tabular}{@{}#1@{}}#2\end{tabular}}

\newcounter{lessoncount}

\newcommand{\lesson}[1]{
\refstepcounter{lessoncount}
\vspace{2pt}
\setlength\fboxrule{0.8pt}
\noindent\fbox{%
\parbox{0.98\linewidth}{%
   \textit{Take-away:} {#1}
}}
\vspace{2pt}
}

\newcommand{\observation}[1]{
\vspace{4pt}
\setlength\fboxrule{0.8pt}
\noindent\fbox{%
\parbox{0.96\linewidth}{%
    {#1}
}}
\vspace{6pt}
}

\definecolor{greencell}{RGB}{146,210,14}
\definecolor{redcell}{RGB}{250,70,11}

\newcommand{\blackthanks}[1]{%
    \begingroup
    \renewcommand{\thefootnote}{\textcolor{black}{\fnsymbol{footnote}}}\unskip
    \thanks{#1}
    \unskip\endgroup
}

%% file: 0-abstract.tex
How can one build a \emph{feature-rich}, \emph{general-purpose}, Rust-based operating system (OS) with a \emph{minimal} and \emph{sound} Trusted Computing Base (TCB) for \emph{memory safety}? Existing Rust-based OSes fall short due to their improper use of \texttt{unsafe} Rust in kernel development. To address this challenge, we propose a novel OS architecture called \emph{framekernel} that realizes Rust's full potential to achieve \emph{intra-kernel privilege separation}, ensuring TCB minimality and soundness. We present \ostd, a streamlined framework for safe Rust OS development, and \aster, a Linux ABI-compatible framekernel OS implemented entirely in safe Rust using \ostd. Supporting over \minor{210} Linux system calls, \aster delivers performance on par with Linux, while maintaining a minimized, memory-safety TCB of only about \minor{14.0\%} of the codebase. These results underscore the practicality and benefits of the framekernel architecture in building safe and efficient OSes.

%% file: 1-intro.tex
\section{Introduction}

Despite three decades of research into defending against memory safety bugs
in operating systems (OSes) written in C,
achieving true memory safety remains elusive. This was starkly demonstrated by the recent CrowdStrike outage \cite{crowd-strike-incident}, where millions of Windows PCs crashed due to an out-of-bounds memory access in a faulty driver. It is estimated that 60-70\% of security vulnerabilities in system software written in C stem from memory safety issues \cite{memory-safety-vulnerabilities}.

In recent years, as the Rust programming language matures and becomes popular, the development of Rust-based, memory-safe OSes has gained momentum. Rust offers memory safety guarantees through innovative language features such as ownership, borrowing, and lifetimes, enabling safe memory management without relying on garbage collection. Many now see Rust as a potential successor to C and C++ as the dominant systems programming language. With the endorsement of Linus Torvalds, the Linux kernel has officially adopted Rust as its second programming language \cite{fisrt-look-at-r4l} and integrated the Rust for Linux \cite{RustLinuxKernel} (RFL) project to facilitate writing "leaf" kernel modules in Rust. Additionally, new OS kernels like Tock \cite{Tock}, RedLeaf \cite{RedLeaf}, and Theseus \cite{Theseus} are built from the ground up using Rust, further demonstrating Rust's potential in this domain.

\begin{table}[tbp]
    \centering
    \footnotesize
    \caption{The \texttt{unsafe} keyword is widely utilized in the crates (or kernel modules) of existing Rust-based OSes. The statistics were derived from an analysis of the latest source code of these OSes at the time of writing.}
    \begin{threeparttable}
    \begin{tabular}{ccccc}
        \toprule
       \textbf{Rust-based OSes} & Linux & Tock & RedLeaf & Theseus \\
       \midrule
        \multirow{2}{11em}{\texttt{Unsafe}-utilizing crates} & 6 / 11\tnote{1}  &  91 / 98  & 36 / 58  & 54 / 171  \\
        & (55\%) &  (93\%) & (62\%) &  (32\%)\\
         \bottomrule
    \end{tabular}
    \begin{tablenotes}[flushleft]
       \item[1] This includes the RFL crate and 10 notable Rust-written kernel modules~\cite{RflWikipedia}.
    \end{tablenotes}
    \end{threeparttable}
    \label{tab:unsafe-util}
\end{table}

While adopting Rust is a significant step toward achieving kernel memory safety, this is insufficient on its own since Rust-based OSes must include unsafe Rust code. The safety of the kinds of low-level controls required by kernel programming cannot be statically verified by the Rust compiler and thus is only allowed by Rust within special code blocks marked by the \texttt{unsafe} keyword. Despite the Rust language team dedicating an entire book \cite{Rustonomicon} to the ``dark arts of unsafe Rust'', developers are still prone to misusing it, and the RustSec Advisory Database has recorded hundreds of bugs stemming from \texttt{unsafe} misuse \cite{Advisory-db}.

To mitigate the risks associated with \texttt{unsafe} in Rust, two widely accepted best practices have emerged~\cite{HowToUseUnsafeRust}: (1) use \texttt{unsafe} sparingly and (2) encapsulate \texttt{unsafe} code within safe abstractions. However, we have found that existing Rust-based OS kernels often fall short of these standards. Unsafe Rust code permeates a significant portion of a Rust-based OS. We observe that \texttt{unsafe}-utilizing crates make up 55\%, 93\%, 62\%, and 32\% of all crates in Linux, Tock, RedLeaf, and Theseus, respectively (as shown in Table~\ref{tab:unsafe-util}). Although kernel developers generally view \texttt{unsafe} as a ``necessary evil'', we question whether such widespread use of \texttt{unsafe} is truly necessary. In particular, we challenge the necessity of \texttt{unsafe} in device drivers (as seen in all existing Rust-based OSes), which account for the majority of the codebase of a mature OS (70\% in Linux \cite{linux-driver-70}).
Details about the pitfalls of \texttt{unsafe} handling in existing Rust OSes will be present in \cref{sec:background}.

\begin{figure}[t]
  \centering
  \includegraphics[width=\linewidth]{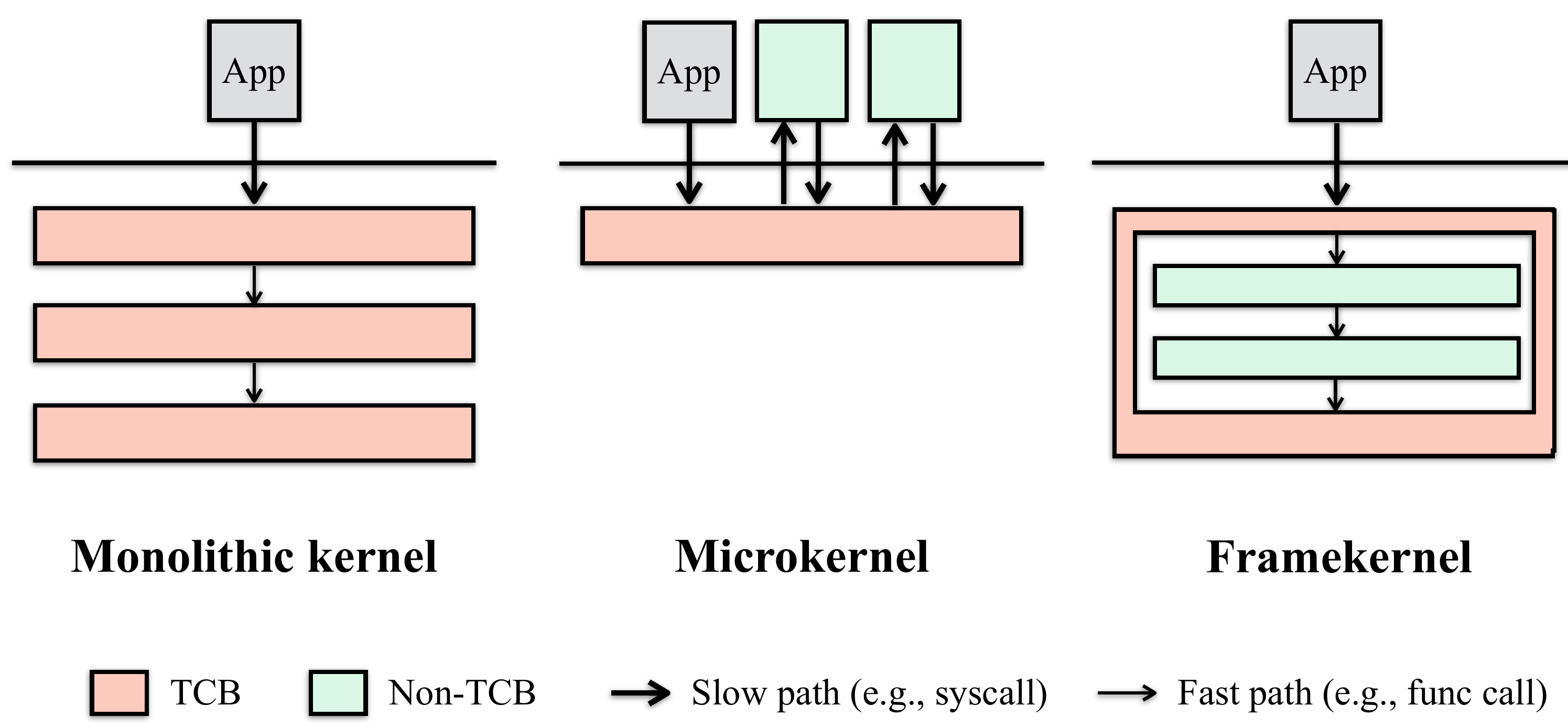}
  \vspace{-0.6cm}
  \caption{\major{A framekernel combines the speed of a monolithic kernel and the security of a microkernel. The memory-safety TCB of a framekernel is reduced to the privileged OS framework (akin to a microkernel) without any communication overheads due to extra hardware-based isolation (similar to a monolithic kernel). We depict the TCB portion of a framekernel as a ``frame'' to highlight the fact all low-level interactions of non-TCB with the hardware (beneath the ``frame'') and the user space (above the ``frame'') is mediated by the TCB.}}
  \label{fig:arch-comparison}
\end{figure}

Given the limitations of prior work, we pose the question: is it possible to build a feature-rich, general-purpose, Rust-based OS kernel almost entirely in safe Rust? We introduce \emph{framekernel}~(\cref{sec:framekernel}), a novel OS architecture designed to achieve a minimal and sound TCB for a Rust-based OS.
\minor{In the framekernel architecture, the entire OS resides in a single address space (as in a monolithic kernel) and is implemented in Rust. The kernel is logically divided into two parts: the \emph{privileged} OS framework (akin to a microkernel) and the \emph{de-privileged} OS services. Only the privileged framework is allowed to use \texttt{unsafe}, while the de-privileged services must be written in safe Rust completely. As the TCB, the privileged framework encapsulates all low-level, hardware-oriented \texttt{unsafe} operations behind safe APIs. Using these safe APIs, the de-privileged OS services can implement all kinds of OS functionalities, including device drivers.
A framekernel minimizes the TCB size without incurring extra overheads due to hardware isolation.
Thus, we claim that a framekernel combines the benefits of both a monolithic kernel and a microkernel (see Figure~\ref{fig:arch-comparison}).}

\minor{We enforce this \emph{language-based, intra-kernel privilege separation} by systematically identifying \emph{sensitive} OS resources, namely those that can be mis-programmed or misused to compromise memory safety -- even with safe Rust. Thus, the design principle of a framekernel is to \emph{keep sensitive OS resources within the privileged framework for soundness, while delegating insensitive OS resources to the de-privileged OS services for minimality}.}

To realize the vision of framekernels,
we develop \ostd~(\cref{sec:ostd}), the privileged OS framework required by a framekernel.
\ostd provides a \emph{small yet expressive} set of safe OS development abstractions,
covering safe user-kernel interactions, safe kernel logic, and safe kernel-peripheral interactions.
Of particular note is the \emph{untyped memory} abstraction,
which addresses the challenge of safely handling \emph{externally-modifiable memory}
(e.g., MMIO or DMA-capable memory) -- a longstanding obstacle in safe driver development.
In addition, we introduce \emph{safe policy injection},
a technique that separates potentially complex "policy" components
(e.g., task schedulers, page allocators, and slab allocators)
from the core "mechanisms" of \ostd,
thus containing the growth of \ostd's complexity over time.
Furthermore, we define a set of key \emph{safety invariants}
that collectively ensure the soundness of \ostd's privilege separation.

To demonstrate the practicality and benefits of framekernels,
we develop \aster~(\cref{sec:aster}), a Linux ABI–compatible framekernel built on \ostd.
\aster implements a rich subset of Linux functionality:  
it supports over \minor{210} Linux system calls,
multiple file systems and socket types,
and a variety of peripherals (e.g., disks and NICs).  
All these features are written in safe Rust, leveraging \ostd's APIs.
\aster has been under development for three years.
The repository of \aster and \ostd is open source~\cite{repo, ae-repo},
containing over \minor{100K lines of Rust code} contributed by more than 50 individuals.

We conduct a thorough evaluation~(\cref{sec:eval}) of \aster and \ostd across three dimensions:
performance, TCB size, and soundness.
\aster delivers performance on par with Linux:
on LMbench, a syscall-intensive microbenchmark,
\aster achieves a mean normalized performance score of \minor{1.08} (relative to Linux, higher is better);
for three I/O-intensive applications, Nginx, Redis, and SQLite,
it delivers a normalized performance of \minor{1.17}, \minor{1.31}, and \minor{0.85}, respectively. 
The framekernel design also yields a lean TCB:
\aster's TCB accounts for just \minor{14.0\%} of its codebase,
versus \minor{43.8\% in Tock, 62.4\% in Theseus, and 66.1\% in RedLeaf}.  
Finally, to strengthen our confidence in \ostd's soundness,
we develop \kernmiri, a retrofitted version of Miri~\cite{miri} for Rust OSes,
which is used to detect potential safety issues in \ostd systematically.

\bheading{Contributions.} In summary, this paper makes the following contributions:
\begin{packeditemize}
\item We propose framekernel, a novel OS architecture that combines the benefits of both monolithic and microkernels by enforcing Rust-based intra-kernel privilege separation.

\item We develop \ostd, a small and sound OS framework to facilitate OS development in safe Rust.

\item We develop \aster, a highly-optimized, Linux ABI-compatible OS based on \ostd.

\item We conduct extensive evaluations to show the performance, TCB size, and safety of both \aster and \ostd.

\end{packeditemize}

%% file: 2-background.tex
\section{Background and Motivation}
\label{sec:background}

\subsection{The Rusty Way to Safety}
\label{subsec:intro_rust}

Rust is an efficient system programming language
that offers strong guarantees for memory, type, and thread safety
without compromising runtime performance,
thanks to its strong type system and unique ownership model.

\bheading{Ownership, borrowing, and lifetime.}
In Rust's ownership model, each value has a single owning variable, and the value's lifetime is tied to the owner's scope~\cite{WhatOwnershipRust}. When the owner goes out of scope, the value is dropped. Ownership can be borrowed through references, subject to lifetime constraints enforced by the compile-time borrow checker.

\bheading{Type system.}
The Rust compiler implements a tailored type system coupled with comprehensive compile-time checks~\cite{TypeCheckingRust, BorrowCheckerRust}. Utilizing type information, the compiler ensures that all accesses occur within valid lifetimes, generates correct memory offsets, and performs bounds checking, thereby guaranteeing both temporal and spatial memory safety.

\bheading{Unsafe Rust.}
To offer additional expressive power, Rust provides the \texttt{unsafe} keyword, enabling programmers to bypass compile-time checks and thus shifting the responsibility for ensuring safety to those using \texttt{unsafe} code blocks~\cite{HowSafeUnsafe}.

\bheading{Undefined behaviors.} \minor{Undefined behaviors (UBs) in Rust refers to operations
that compromise the language's correctness and safety guarantees,
including memory safety, thread safety, and type safety.
The Rust Reference book enumerates common UBs~\cite{BehaviorConsideredUndefined}, e.g.,
data races,
memory access based on dangling or misaligned pointers,
out-of-bound memory access,
breaking the pointer aliasing rules,
and mutating immutable bytes.
However,
this list is neither comprehensive nor precise
due to Rust's lack of a formalized specification.
The Rust language team maintains an official UB detection tool called Miri~\cite{miri},
which, as part of the Rust toolchain,
serves as a de-facto executable standard for UBs.
Although Miri can effectively detect UBs in standard Rust applications,
Miri is not applicable to Rust-based OSes 
as its design does not take into account the needs of low-level system programming,
particularly in areas of memory management and hardware interaction.
This work aims at systematic identification and prevention of UBs
in the context of Rust OSes.}

\subsection{Rustification of Mainstream OSes}

\minor{The efficiency and safety of Rust make it an attractive choice for developing OSes, leading to several mainstream OSes~\cite{RustLinuxKernel, RustForWindows, RustForAndroid} integrating Rust into their codebases. The Rust for Linux (RFL) project~\cite{RustLinuxKernel} stands out as a prominent example.}

RFL aims at establishing Rust as the second official language of the Linux kernel, allowing developers to write ``leaf'' kernel modules in safe Rust. Officially merged into Linux in 2022, RFL has laid the groundwork for Rust integration by 2024~\cite{RFLSurvey}. Notably, three Rust-written device drivers, albeit basic, have already been added to the kernel tree~\cite{empirialRustLinux}.

Despite this progress, the unsafe nature of Linux remains.
By design, RFL has to offer safe Rust abstractions over Linux's extensive legacy C APIs,
resulting in substantial use of \texttt{unsafe} Rust.
A recent study~\cite{empirialRustLinux} notes that
RFL already has 19K lines of Rust code upstream,
with an additional 112K lines staged for upstream inclusion.
A great portion of RFL is devoted to unsafe Rust code that interacts with the legacy C APIs.
As RFL expands to cover more subsystems and their C APIs,
we expect its codebase to grow to hundreds of thousands of lines. 
Therefore, the TCB size of RFL is substantial,
even not considering Linux's huge C core and the countless C kernel modules around it.

\begin{framed}
\textbf{Lesson Learned:} Constructing safe Rust abstractions on a legacy monolithic kernel inevitably requires a substantial use of \texttt{unsafe} Rust.
\end{framed}

\minor{In addition to the inflated TCB size,
the burden of a huge legacy codebase - its status quo and established philosophy -
constrains the effectiveness of Rust and the soundness of RFL.
This constraint leaves known soundness vulnerabilities unresolved in RFL.
For example, 
Rust explicitly excludes memory leaks from its memory safety guarantees,
permitting objects to be forgotten.
However, it was discovered that forgetting RFL's mutex guards can trigger use-after-free vulnerabilities~\cite{rflUAF}.
This vulnerability arises from conflicting API contracts between C and Rust
regarding mutex unlocking obligations,
so neither side is willing or capable to fix it.
Another example of a soundness issue stems from Linux's (and by extension RFL's) tolerance for sleeping in atomic contexts
such as spinlock or RCU-lock held regions.
Sleep-in-atomic-context bugs may cause data races in RCU-protected memory accesses,
undermining Rust API safety guarantees~\cite{Klint}. 
These unresolved soundness issues reflect the Linux community's traditional value of 
"pragmatism over safety"~\cite{Klint},
creating fundamental tensions with Rust's security-first paradigm.}

\begin{framed}
\textbf{Lesson Learned:} Achieving sound memory safety requires a clean-slate OS that prioritizes safety above all else.
\end{framed}

\subsection{Clean-Slate Rust OSes}

\begin{table}[t]
\centering
\scriptsize
\caption{Representative \texttt{unsafe} Rust code pattern in drivers}
\label{tab:typical-unsafe-patterns}
\begin{tabular}{cll}
\toprule
~ & \makecell{Resource \textit{access}\\ within \texttt{unsafe} blocks} & \makecell{Resource \textit{acquisition} \\ within \texttt{unsafe} blocks} \\
\midrule
\textbf{Tock} & \begin{lstlisting}
// chips/nrf52:
buf[idx] = *byte_ptr
\end{lstlisting} & \begin{lstlisting}
// chips/nrf52:
StaticRef::new(ptr)
\end{lstlisting} \\
\textbf{RedLeaf} & \begin{lstlisting}
// lib/devices/ixgbe:
ptr::read_volatile(addr)
\end{lstlisting} & \begin{lstlisting}
// lib/devices/tpm:
MmioAddr::new(base,len)
\end{lstlisting} \\
\textbf{Theseus} & \begin{lstlisting}
// kernel/pci:
pci_port.write(val)
\end{lstlisting} & \begin{lstlisting}
// kernel/ixgbe:
Box::from_raw(ptr)
\end{lstlisting} \\

\bottomrule
\end{tabular}
\end{table}

Clean-slate Rust OSes like Tock~\cite{Tock}, RedLeaf~\cite{RedLeaf}, and Theseus~\cite{Theseus} strive to fully leverage Rust’s features to improve OS safety and reliability. However, they face a notable limitation: inadequate support for safe driver development. As shown in Table~\ref{tab:typical-unsafe-patterns}, device drivers in these systems frequently rely on \texttt{unsafe} code to manage low-level resources, such as raw data buffers, MMIO, I/O ports, and DMA regions. Given that drivers typically constitute the largest portion of an OS codebase, extensive use of \texttt{unsafe} code significantly heightens the risk of memory safety vulnerabilities.

\begin{framed}
\textbf{Lesson Learned:} Safe driver development requires safe abstractions for acquiring and accessing low-level system resources.
\end{framed}

Among the three Rust OSes, Theseus minimizes \texttt{unsafe} code through its \texttt{MappedPages} abstraction, which represents the \emph{exclusive} ownership of a virtually-contiguous memory region backed by some \emph{exclusively}-owned physical frames. This exclusiveness alone (seemingly) preserves the safety of Rust references (e.g., \texttt{\&T}) borrowed from that range.
However, this design overlooks \emph{sensitive} or \emph{externally-modifiable} memory. For example, using \texttt{MappedPages} to modify the sensitive memory of Local APIC may cause unpredictable CPU behavior (e.g., sending an IPI that resets a CPU). Similarly, Theseus drivers create Rust references (\texttt{\&T}) to MMIO device memory, an unsound practice since such references assume no external modifications, which hardware may violate.

\begin{framed}
\textbf{Lesson Learned:} Rust OSes must develop safe abstractions for memory that is subject to external modifications, such as hardware, user programs, or DMA.
\end{framed}

Finally, while language-level UBs are effectively addressed by Rust's safety guarantees, UBs stemming from execution environments or CPU architectures remain unresolved in existing Rust-based OSes. For example, a malicious device could corrupt kernel memory via DMA~\cite{thunderclap} or spoof interrupts to manipulate CPU trap handlers~\cite{interruptSpoofing}. Similarly, a stack overflow could compromise the execution environment—an issue beyond the detection capabilities of safe Rust.

\begin{framed}
\textbf{Lesson Learned:} Rust OSes should safeguard against UBs not only at the language level but also at the architectural and environmental levels.
\end{framed}

%% file: 3-framekernel.tex
\section{Framekernel Architecture}
\label{sec:framekernel}

We introduce \textit{framekernel}, a novel OS architecture that combines the benefits of both monolithic kernels and microkernels by fully leveraging the modern safe system language of Rust. In a framekernel OS, the entire OS kernel resides within a single address space, akin to a monolithic kernel, and is implemented in Rust. The kernel is logically divided into two parts: the \textit{privileged OS framework} (similar to a microkernel) and the \textit{de-privileged OS services}. Only the privileged framework may use Rust's \texttt{unsafe} features, while the de-privileged services are built entirely in safe Rust. The privileged framework encapsulates low-level, hardware-oriented unsafe operations into safe APIs, using which the de-privileged services implement most OS functionalities, including drivers, in safe Rust. Like a monolithic kernel, all components within a framekernel communicate efficiently (e.g., via function calls or shared memory). By restricting the privileged framework to a minimal set of functionalities, framekernels, like microkernels, reduce the size of the TCB, thereby enhancing safety, security, and reliability.

\bheading{Design philosophy.}
At the heart of the framekernel architecture is \textit{intra-kernel privilege separation}: although OS services operate in privileged CPU mode, their behaviors are constrained by safe Rust and the privileged framework, maintaining their de-privileged status. Only flaws within the privileged framework can jeopardize the kernel's memory safety. This intra-kernel privilege separation must uphold two key properties: soundness and minimality.

\begin{framed}
\textbf{Soundness:} The OS framework guarantees the absence of UBs under all circumstances, irrespective of interactions with OS services, user code, or peripheral devices.
\end{framed}

This property aims to prevent all forms of UBs, which may originate at three levels (from high to low). First, \textbf{language-level UBs} in Rust are described in \cref{subsec:intro_rust}. Second, \textbf{environment-level UBs} arise when the code, stack, or heap is corrupted. Programming languages have implicit assumptions about their execution environments; UB occurs if those assumptions are compromised. Third, \textbf{architecture-level UBs} stem from incorrect use of CPU or hardware devices, such as improperly saving/restoring CPU registers, misconfiguring page tables, or allowing peripheral DMA to corrupt memory.

\begin{framed}
\textbf{Minimality:} A component is tolerated inside the OS framework only if moving it outside would prevent the implementation of OS services' required functionality or compromise soundness of the framework.
\end{framed}

This property focuses on minimizing the principal component of the \emph{runtime TCB} responsible for kernel memory safety. Naturally, the whole TCB extends beyond the OS framework itself, also including the Rust toolchain, the Rust core libraries, bootloader, and firmware involved in loading a framekernel, as well as the CPU plus some core devices (e.g., interrupt controller and IOMMU). \major{The software and hardware outside the TCB, including the safe OS services, user-space programs, and peripheral devices (e.g., disks, NICs, and GPUs), are not trusted.}

\begin{figure}[t]
  \centering
  \includegraphics[width=\linewidth]{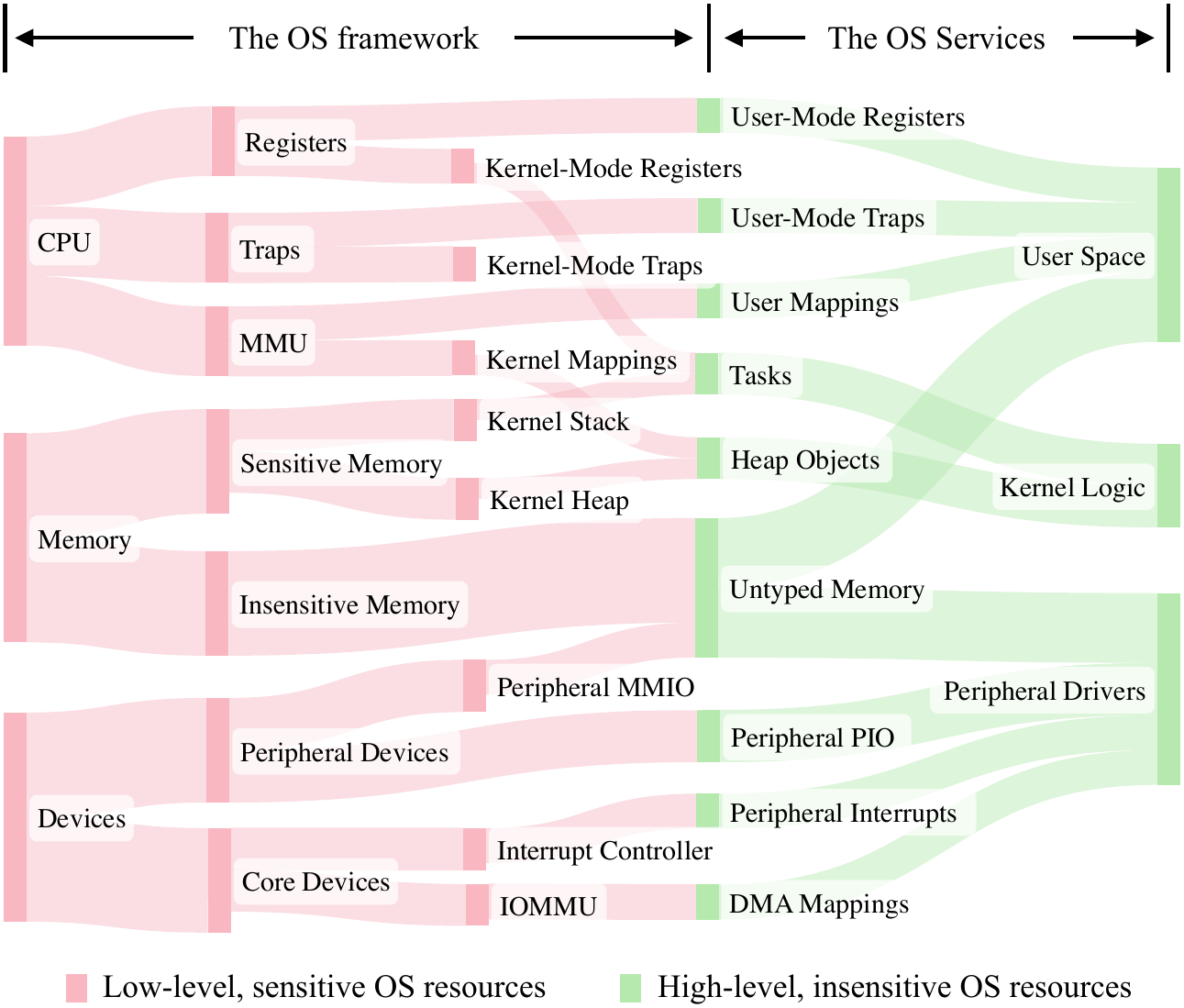}
  \vspace{-0.6cm}
  \caption{A blueprint for framekernel construction in a \emph{resource}-centric view. The nodes represent OS resources, while the edges depict classification and encapsulation. The concept of untyped memory will be introduced in \cref{subsec:frame}.}
  \label{fig:framekernel_blueprint}
\end{figure}

\bheading{Towards effective privilege separation.}
Realizing intra-kernel privilege separation prompts a key question: \emph{What must stay inside the framework (for soundness), and what can be moved outside (for minimality)?}

To answer this, we view framekernel construction from a \emph{resource}-centric perspective, as shown in Figure~\ref{fig:framekernel_blueprint}. Fundamentally, an OS manages three classes of resources: CPU, memory, and devices.  At first glance, these resources are all \textbf{sensitive} because the kernel code could misuse them (even in safe Rust) and break memory safety. However, deeper inspection shows that some subsets of these resources are actually \textbf{insensitive}. Thus, we \emph{must keep sensitive resources inside the framework (for soundness) and should move insensitive ones outside (for minimality).}

All three classes of fundamental resources can be split into sensitive and insensitive subsets. \textbf{CPU} resources involve kernel-mode and user-mode control; the former is considered sensitive, while the latter (e.g., user-mode registers, user-mode traps) is insensitive because it cannot directly undermine kernel state. Furthermore, the user virtual address space is insensitive as manipulating the user virtual memory does not affect kernel memory safety. \textbf{Memory} resources include sensitive memory used for the kernel's code, stack, heap, and page tables, whereas memory dedicated to untrusted user processes or devices is insensitive because kernel safety does not depend on its integrity. \textbf{Devices} present a similar division, as core devices (e.g., APIC, IOMMU) are sensitive due to their machine-wide control, but peripheral devices (e.g., NICs, GPUs) are generally insensitive; a misconfigured core device can compromise the entire kernel, but failures in a peripheral device are typically confined to that device itself.

These observations inform a blueprint for framekernel construction (Figure~\ref{fig:framekernel_blueprint}). \minor{The blueprint \textit{classifies} OS resources into sensitive and insensitive ones at a fine granularity, with low-level, sensitive ones \emph{hidden} inside the privileged OS framework or \emph{encapsulated} into high-level, insensitive ones.} The net result is that the privileged OS framework only exposes to the de-privileged OS services a small set of insensitive resources that is sufficient to support the three primary needs of safe OS development: (1) safe user-kernel interactions, (2) safe kernel logic, and (3) safe kernel-peripheral interactions. This blueprint guides the design of \ostd.

%% file: 4-ostd.tex
\section{\ostd}
\label{sec:ostd}

In this section, we present \ostd, which is our implementation of the framework required by a framekernel. First, we provide an overview of \ostd's APIs. Then, we describe how \ostd manages physical memory pages (frames), a key foundation of \ostd's soundness. Next, we define the key safety invariants for \ostd to achieve the language-based, intra-kernel privilege separation. Lastly, we introduce the technique of safe policy injection, which confines the complexity growth of \ostd as it evolves.

\input{4.1-apis}
\input{4.2-frame}
\input{4.3-seperation}
\input{4.4-injection}

%% file: 4.1-apis.tex
\subsection{Expressive APIs}

\begin{figure}[t]
  \centering
  \includegraphics[width=0.9\linewidth]{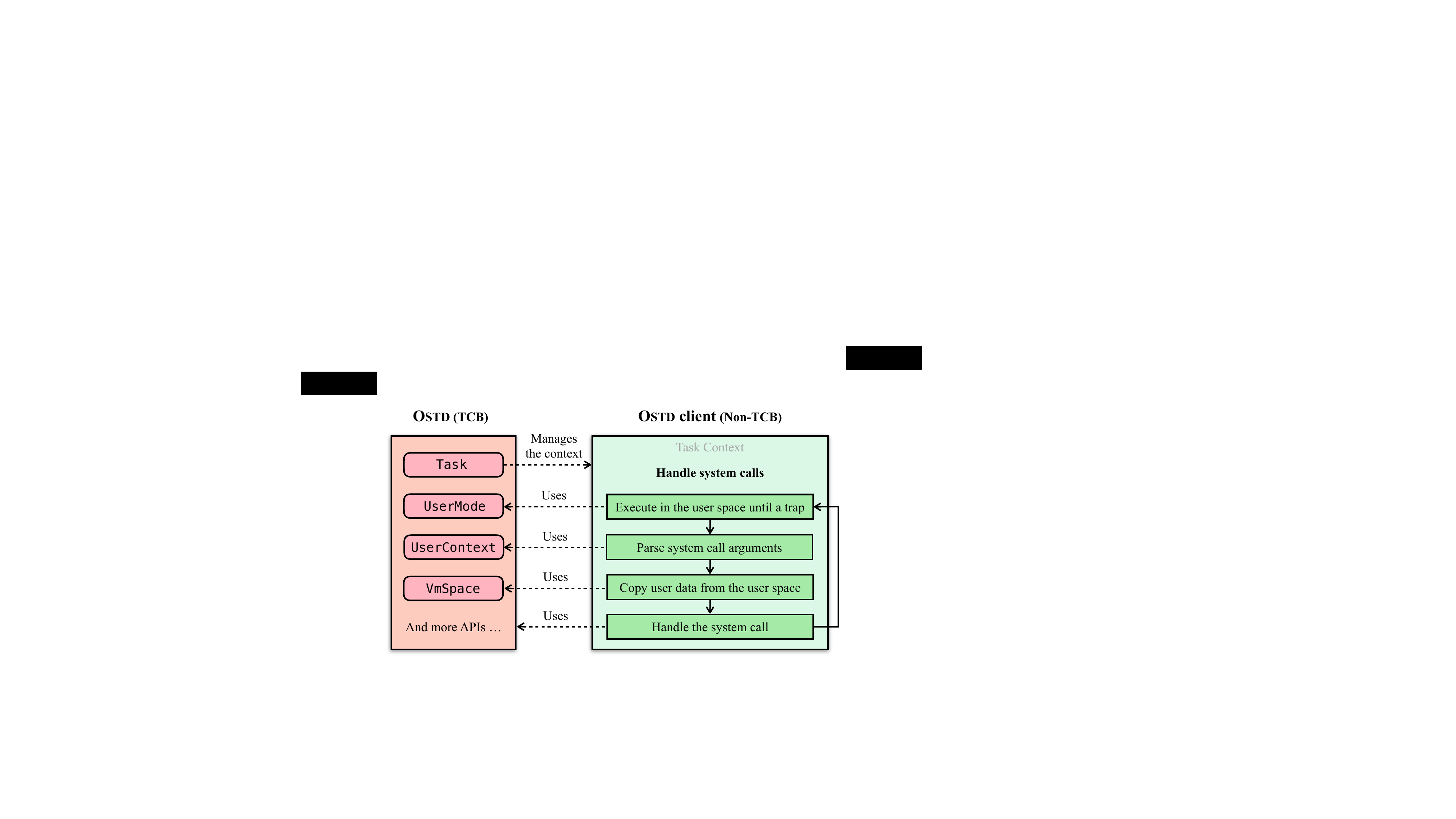}
  \vspace{-0.1cm}
  \caption{An example of \ostd APIs: system call handling}
  \label{fig:ostd_example}
\end{figure}

\ostd provides a small set of expressive abstractions to meet the needs of safe OS development. These abstractions are implemented as \ostd APIs, which are utilized by \ostd clients -- safe kernel code that interacts with the \ostd. For \textbf{safe user-kernel interactions}, it allows an \ostd client to jump into the user space and execute until a trap occurs (\texttt{UserMode}), manipulate user-mode CPU registers (\texttt{UserContext}), and manage user address spaces (\texttt{VmSpace}). For \textbf{safe kernel logic}, it includes synchronization primitives (e.g., \texttt{SpinLock}, \texttt{Rcu}, \texttt{Mutex}, \texttt{WaitQueue}, and \texttt{CpuLocal}) and efficient data collection types (e.g., \texttt{LinkedList} and \texttt{RbTree}). For \textbf{safe kernel-peripheral interactions}, it offers mechanisms to register interrupt handlers (\texttt{IrqLine}), perform MMIO and PIO (\texttt{IoMem} and \texttt{IoPort}), and create coherent or streaming DMA mappings (\texttt{DmaCoherent} and \texttt{DmaStream}). All three of these OS development aspects may allocate and access physical pages, referred to as \textbf{frames}, via \texttt{Frame} (one frame) or \texttt{Segment} (multiple contiguous frames).

We illustrate a simplified workflow of how system calls may be handled safely with \ostd in Figure~\ref{fig:ostd_example}. Another example on how device data is requested with \ostd is shown in Figure~\ref{fig:ostd_example2} in Appendix~\ref{appendix:ostd_example2}. For a more concrete example, see our sample project ``Write a Hello World OS Kernel in $\sim$100 Lines of Safe Rust with \ostd''~\cite{ExampleWritingKernel}. 

%% file: 4.2-frame.tex
\subsection{Frame Management}
\label{subsec:frame}

\ostd employs a \textbf{frame metadata system} to track the state of each page frame, including a reference count and a customizable metadata field. All per-frame metadata are stored in a large static array, which is allocated and initialized in an early bootstrap phase. The reference count equals the total number of \texttt{Frame} and \texttt{Segment} instances that refer to that frame, enabling \ostd to manage frame lifetimes correctly.

\begin{inv}
\label{inv:frame_unused}
Any newly-allocated \texttt{Frame} or \texttt{Segment} originates from currently unused memory.
\end{inv}

\ostd further supports custom per-frame metadata by taking a type parameter \texttt{M} in \texttt{Frame<M>} (or \texttt{Segment<M>}). The client can attach a metadata object of \texttt{M} to a \texttt{Frame<M>} object, enabling features like a page cache to efficiently maintain some extra per-frame states (e.g., the status of data synchronization between memory and disk). This mechanism is also used in page allocators and slab allocators (see~\cref{subsec:safe-policy-injection}).

Both \texttt{Frame} and \texttt{Segment} can represent either sensitive or insensitive memory. Internally, \ostd uses them to allocate sensitive memory for kernel resources such as page tables, stacks, or slabs. Externally, an \ostd client can request a special form of insensitive memory called \textbf{untyped memory} with \texttt{UFrame<M> = Frame<M: AnyUFrameMeta>} or \texttt{USegment<M> = Segment<M: AnyUFrameMeta>}, where the trait \texttt{AnyUFrameMeta} marks metadata types suitable for untyped memory usage.

Untyped memory deals with the fact that externally modifiable memory (e.g., user-mapped or DMA-capable memory) cannot uphold the strong guarantees of Rust references or the type safety of arbitrary Rust types. As such, we design \texttt{UFrame} (or \texttt{USegment}) to have a \textbf{read-write style interface}~\cite{untyped-mem} that only permits copying \textbf{plain old data} (POD)~\cite{cppPOD} from or into it. A POD type (e.g., \texttt{u32}) can hold a value of any bit pattern, without invalidating any Rust invariants.

%% file: 4.3-seperation.tex
\subsection{Privilege Separation}

Intra-kernel privilege separation requires \emph{preventing all sensitive resources from being tampered with by non-TCB entities}, including safe clients, user programs, and peripheral devices. \ostd achieves this by enforcing the following key invariants.

\begin{inv}
Kernel-mode CPU states cannot be tampered with by \ostd clients.
\end{inv}

To expose only user-mode CPU operations to clients, \ostd provides \texttt{UserMode}, \texttt{UserContext}, and \texttt{VmSpace}. The first two handle traps and registers at user privilege level, while the third manages user-mode virtual memory. Kernel-mode registers are hidden, and only safe portion of user-mode CPU registers are accessible. For instance, on x86, \texttt{UserContext} exposes only the non-sensitive subset of \texttt{RFLAGS}, excluding bits like \texttt{IF} or \texttt{IOPL} that control interrupts or I/O privilege.

\begin{inv}
Kernel-mode CPU states cannot be tampered with by peripheral devices.
\end{inv}

On x86 hardware, devices can spoof exceptions, traps, or inter-processor interrupts~\cite{interruptSpoofing}. To protect against such attacks, \ostd configures the IOMMU to enable interrupt remapping, preventing rogue devices from influencing kernel control flow.

\begin{inv}
Sensitive memory cannot be tampered with by \ostd clients.
\end{inv}

All \texttt{Frame}s or \texttt{Segment}s allocated by \ostd clients originate from insensitive physical memory. Sensitive pages, such as those used for kernel code, stacks, or page tables, remain fully within \ostd's control.

Each \texttt{Task}'s stack includes a guard page to detect stack overflows. If the stack pointer touches the guard page, a page fault is triggered, preventing further execution and thwarting potential malicious behavior. Additionally, the \ostd-based OS enforces a compile-time stack usage analysis, ensuring each function's stack frame remains smaller than the guard page. This prevents malicious attempts to bypass the guard page and exploit stack overflows.

\begin{inv}
Sensitive memory cannot be tampered with by user programs.
\end{inv}

User-mode mappings are created through \texttt{VmSpace}, which can only take \texttt{UFrame} or \texttt{USegment} as inputs. This design rules out exposing sensitive memory to user space because untyped memory are, by definition, insensitive.

\begin{inv}
Sensitive memory (including I/O memory) cannot be tampered with by peripheral devices.
\end{inv}

\ostd leverages IOMMU to prevent peripheral devices from writing to unauthorized physical regions. Initially, no part of physical memory is DMA-accessible. Drivers can create DMA mappings (\texttt{DmaStream} or \texttt{DmaCoherent}) only over untyped memory (\texttt{UFrame} or \texttt{USegment}), so sensitive regions stay protected.

\begin{inv}
Sensitive I/O memory or ports cannot be tampered with by \ostd clients.
\end{inv}

Clients interact with MMIO and PIO through \texttt{IoMem} and \texttt{IoPort}. \ostd uses information from the architecture, firmware (e.g., ACPI tables on x86), and core device drivers to label ranges of I/O memory and ports as either sensitive or insensitive. \texttt{IoMem} and \texttt{IoPort} can only be instantiated for insensitive regions, preventing accidental or malicious misuse of sensitive I/O registers.

%% file: 4.4-injection.tex
\subsection{Safe Policy Injection}
\label{subsec:safe-policy-injection}

One's confidence in the correctness of any safe abstraction ultimately hinges on the size and complexity of its TCB. In this section, we explain how to contain \ostd---even if its implementation adopts increasingly sophisticated strategies and policies over time.

Specifically, \ostd consists of the following components, each of which makes decisions based on particular strategies or policies: (1) \textbf{Task scheduler} determines which task to run next, (2) \textbf{Frame allocator} decides how to allocate large chunks of memory, and (3) \textbf{Slab allocator} decides how to allocate smaller chunks of memory.
These components can grow significantly when equipped with advanced features. For instance, their counterparts in Linux have grown in size and sophistication over the years, as summarized in Table~\ref{tab:linux_complexities}.

\begin{table}[]
\centering
\footnotesize
\caption{Increased complexities in some Linux components}
\label{tab:linux_complexities}
\begin{tabular}{llrl}
\toprule
\begin{tabular}[c]{@{}l@{}}Linux\\ Components\end{tabular} &
  \multicolumn{1}{c}{\begin{tabular}[c]{@{}c@{}}Early Version\\ (2.1.23, 1997)\end{tabular}} &
  \multicolumn{2}{c}{\begin{tabular}[c]{@{}c@{}}Latest Version\\ (6.12.0, 2024)\end{tabular}} \\
\midrule
Task scheduler  & 1.6 KLoC & 27.2 KLoC & \textbf{$17\times$} \\
Slab allocator  & 1.6 KLoC & 8.7 KLoC  & \textbf{$6\times$}  \\
Frame allocator & 1.2 KLoC & 7.1 KLoC  & \textbf{$6\times$}  \\
\bottomrule
\end{tabular}
\end{table}

Intuitively, a TCB should contain only mechanisms rather than policies for minimality. We therefore propose \textbf{safe policy injection}, a technique that removes complex policies from a Rust TCB without compromising functionality, efficiency, or soundness. To apply the technique, a developer identifies components inside the TCB that might use complex strategies or policies, then determines whether a complete policy implementation can be written in safe Rust. If it can, the developer designs abstractions for acceptable policies plus APIs to register those policies, a process we refer to as ``injection''.

Although the idea of safe policy injection appears simple, ensuring soundness can be challenging because these policies affect the behaviors of TCB code. For example, a safe-but-buggy scheduler could inadvertently schedule the same task on two CPUs at once -- a catastrophic mistake for memory safety. Similarly, the choice of which memory page or slot to allocate next directly affects memory safety. All mainstream OSes include frame and slab allocators within their TCBs for precisely these reasons. The rest of this subsection describes how we overcome such challenges.

\subsubsection{Task Scheduler}

\begin{table}[t]
\centering
\footnotesize
\caption{Select APIs for task scheduler injection. \texttt{Scheduler} represents a task scheduler, and each CPU's local run queue is represented by \texttt{RunQueue}.}
\label{tab:scheduler_api}
\begin{tabular}{p{5.0cm}p{2.7cm}}
\toprule
\textbf{APIs} & \textbf{Descriptions} \\
\midrule
\texttt{Scheduler::enqueue(\&self, task)} & Enqueue a task\\
\texttt{Scheduler::local\_rq\_with(\&self, closure)} & Access to the local run queue with a closure \\
\texttt{RunQueue::update\_curr(\&mut self)} & Update current task\\
\texttt{RunQueue::pick\_next(\&mut self)} & Pick the next task \\
\texttt{RunQueue::dequeue\_curr(\&mut self)} & Remove the current task \\
\bottomrule
\end{tabular}
\end{table}

Our goal is to support advanced schedulers, such as Linux's Completely Fair Scheduler (CFS)~\cite{CFSSchedulerLinux}, atop \ostd. We introduce to \ostd two new traits: \texttt{Scheduler} and \texttt{RunQueue}. Their APIs are summarized in Table~\ref{tab:scheduler_api}. A type implementing \texttt{Scheduler} should be registered once at an early stage of kernel initialization (when no tasks exist). Whenever a task becomes runnable (e.g., is spawned or woken), \ostd hands it over to the scheduler via the \texttt{enqueue} method.

We require schedulers to be SMP-friendly by keeping per-CPU run queues, accessible through \texttt{local\_rq\_with}. To replace the current task, \ostd calls \texttt{pick\_next}, and if the current task becomes unrunnable (e.g., it sleeps), \texttt{dequeue\_curr} removes it from the queue. At each scheduling event (e.g., sleep, yield, or timer tick), \ostd invokes \texttt{update\_curr} to notify the scheduler, which can then update the scheduling information about the current task.

For efficiency, the scheduler API takes ownership of runnable tasks (\texttt{Arc<Task>}). Because the task objects are clonable, they can be stored in advanced data structures (e.g., red-black trees) or moved between CPU queues for load balancing. Each task may also carry custom data (\texttt{Box<dyn Any>}) for storing scheduling attributes, making these attributes cheaply accessible.

The client-provided \texttt{pick\_next} method could return an invalid task. In particular, returning a task that is already running on another CPU may lead to severe consequences, as running one task on two CPUs corrupts its stack. We must thus preserve this key invariant:

\begin{inv}
A \texttt{Task} runs on at most one CPU at any given time.
\end{inv}

To enforce this, \ostd gives each \texttt{Task} a private flag, \texttt{is\_running}, which is checked and set prior to a context switch. After switching, the previous task will have its flag cleared.

\subsubsection{Frame Allocator}
\label{subsubsec:frame_allocator_injection}

\begin{table}[t]
\centering
\footnotesize
\caption{Select APIs for frame allocator injection. The \texttt{FrameAlloc} trait abstracts any injectable frame allocator.}
\label{tab:frame_alloc_api}
\begin{tabular}{p{5.5cm}p{2.2cm}}
\toprule
\textbf{APIs} & \textbf{Descriptions} \\
\midrule
\texttt{FrameAlloc::alloc(\&self, layout)} & Allocate frames\\
\texttt{FrameAlloc::dealloc(\&self, addr, size)} & Deallocate frames\\
\texttt{FrameAlloc::add\_free\_memory(\&self,\newline addr, size)} & Add a range of \newline usable frames\\
\bottomrule
\end{tabular}
\end{table}

\minor{An injectable frame allocator is abstracted by the \texttt{FrameAlloc} trait,
whose APIs are shown in Table~\ref{tab:frame_alloc_api}. 
During the initialization phase,
\ostd passes the information of all usable physical memory to the injected frame allocator (\texttt{FrameAlloc::add\_free\_memory}).
Subsequently, whenever an \ostd client requests a new \texttt{Frame} (or \texttt{Segment}),
\ostd redirects the request to the injected allocator (\texttt{FrameAlloc::alloc}),
which returns the address of the allocated memory.
Conversely, when a \texttt{Frame} (or \texttt{Segment})
is dropped and its reference count is reduced to zero,
the underlying frames will be returned to the injected allocator
(\texttt{FrameAlloc::dealloc}).}

\minor{A \texttt{Frame} (or \texttt{Segment}) must be
created out of a \emph{valid} and \emph{unused} range of physical memory.
But this safety pre-condition may be violated by an injected frame allocator
due to potential logical bugs.
To guard against such bugs,
\ostd \emph{only} turns memory ranges obtained
from the injected allocator into \texttt{Frame}s through a safe constructor method, \texttt{Frame::from\_unused(addr, size)},
which enforces Inv. \ref{inv:frame_unused}
by leveraging the frame metadata system (\cref{subsec:frame}).}

\subsubsection{Slab Allocator}

\begin{table}[t]
\footnotesize
\centering
\caption{Select APIs for slab allocator injection. A \texttt{Slab} represents one or more memory pages arranged into fixed-size slots, with free slots represented by \texttt{HeapSlot}.}
\label{tab:slab-apis}
\begin{tabular}{p{4.2cm}p{3.0cm}}
\toprule
\textbf{APIs} & \textbf{Descriptions} \\
\midrule
\texttt{Slab::new()} & Create a new slab \\
\texttt{Slab::alloc(\&self)} & Allocate a new \texttt{HeapSlot} \\
\texttt{Slab::dealloc(\&self, slot)} & Recycle a \texttt{HeapSlot} \\
\texttt{HeapSlot::into\_box(self, val)} & Convert into a heap object\\
\bottomrule
\end{tabular}
\end{table}

We now describe how \ostd supports a custom slab allocator~\cite{slabSingle, slabMany} and its injection as the heap allocator. A \textbf{slab} is one or more contiguous pages partitioned into an array of fixed-size \textbf{slots}~\cite{slabSingle}. Each slot can hold a single object of a particular type or size. A \textbf{slab cache} pools these slots for rapid allocation and release. Typically, a per-CPU free list tracks empty slots, refilled from a global pool of slabs when needed. If all slabs are full, new slabs are allocated from free pages. Conversely, the allocator can free unused slabs to reclaim memory. An OS typically manages multiple slab caches for different slot sizes, all governed by a \textbf{slab allocator}.

We introduce two new abstractions in \ostd: \texttt{Slab} and \minor{\texttt{HeapSlot}}, whose APIs are summarized in Table~\ref{tab:slab-apis}. These APIs perform type conversions critical to memory safety: from an unused memory page to a slab (\texttt{Slab::new}), then to a free slot (\texttt{Slab::alloc}), and ultimately to a heap object (\texttt{HeapSlot::into\_box}). With these abstractions, developers can implement slab caches and the slab allocator in \aster using only safe Rust.

These abstractions maintain some key invariants. For example, a \texttt{Slab} owns its underlying pages, so dropping the \texttt{Slab} should free those pages -- unless some slots are still occupied by active objects. To avoid use-after-free, each \texttt{Slab} tracks the number of active \minor{\texttt{HeapSlot}}s it spawned. A panic is triggered if a \texttt{Slab} is dropped while any slot remains active:

\begin{inv}
A \minor{\texttt{HeapSlot}} or any object derived from it must not outlive its parent \texttt{Slab}.
\end{inv}

\minor{Additionally, when \texttt{HeapSlot::into\_box} is called, the method checks cheaply whether the slot's address and size can fit a requested object of type \texttt{T}:}

\begin{inv}
An object is created from a \texttt{HeapSlot} only if the slot meets the object's size and alignment requirements.
\end{inv}

\minor{In practice, many kernel components rely on a global heap rather than creating their own slab caches. \ostd allows injecting a slab-based, global heap allocator, which can dispatch each heap allocation to the appropriate slab cache.}

Overall, the safe policy injection technique allows \ostd to enjoy the performance advantage of advanced implementations without bloating its TCB size.

%% file: 5-aster.tex
\section{\aster}
\label{sec:aster}

\begin{figure}[t]
  \centering
  \includegraphics[width=\linewidth]{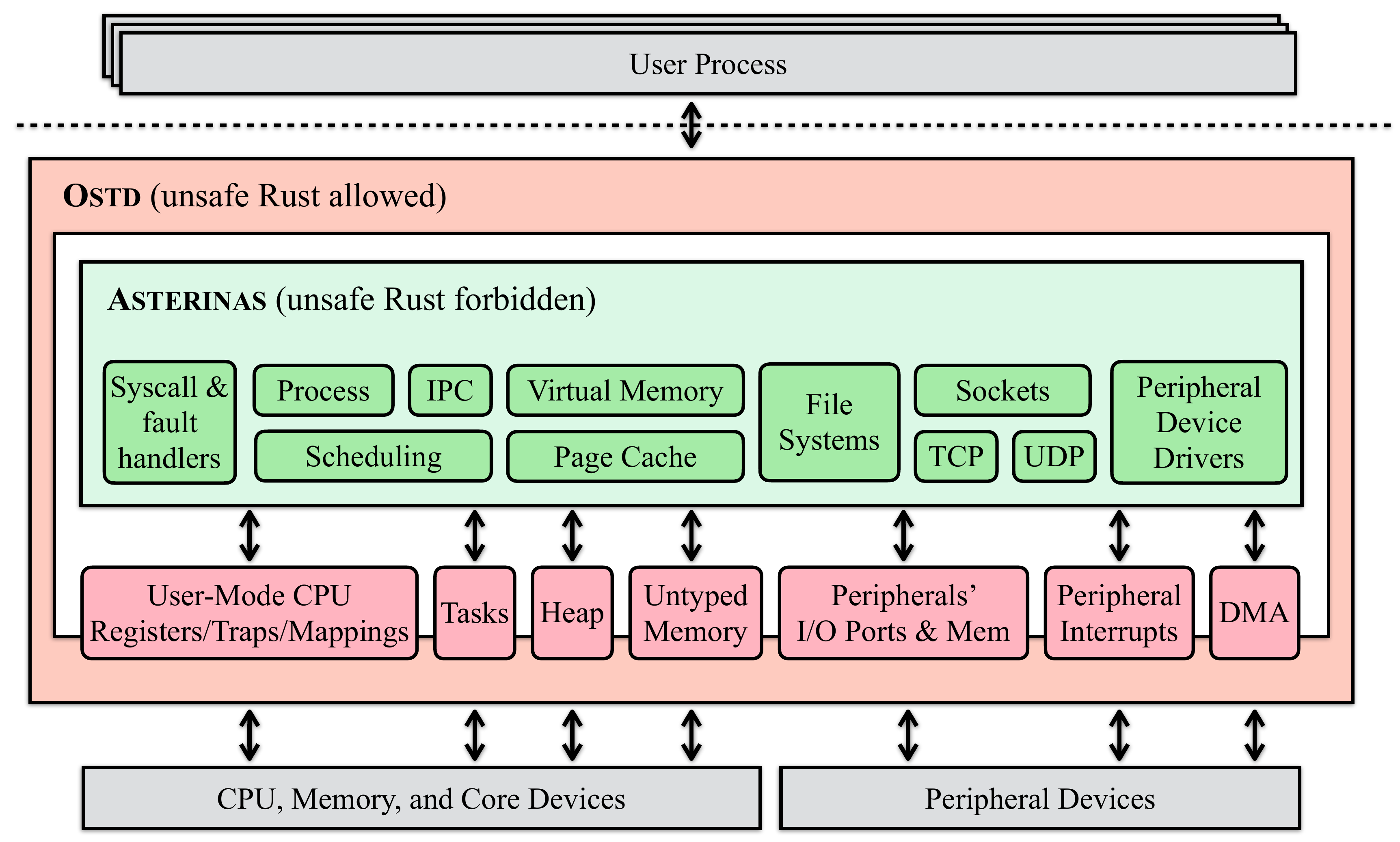}
  \vspace{-0.6cm}
  \caption{\major{An overview of Asterinas.}}
  \label{fig:asterinas-arch}
\end{figure}

We develop \aster (see Figure~\ref{fig:asterinas-arch}), a framekernel-based OS built on top of \ostd. \aster implements a substantial subset of Linux features, including virtual memory, user processes, preemptive scheduling, IPC, a page cache, a virtual file system, and sockets, providing over 210 system calls. It supports various file systems (e.g., Ext2, exFAT32, OverlayFS, RamFS, ProcFS, and SysFS), socket types (e.g., TCP, UDP, Unix, and Netlink), and devices (e.g., Virtio Block, Virtio Network, Virtio Vsock, USB controller, and USB HID). Two CPU architectures are supported: x86-64 (tier-1) and RISC-V (tier-2). All the functionality is written in safe Rust by leveraging \ostd's APIs. \aster has been under development for three years (see Figure~\ref{fig:TCB}). The repository of \aster and \ostd is open-sourced~\cite{repo, ae-repo}, contains over 100K lines of Rust contributed by over 50 individuals.

\bheading{Towards smaller TCB.} In pursuit of a smaller TCB, \aster (rather than \ostd) implements much of the OS infrastructure that Linux considers part of its core. For instance, \aster manages all interrupt bottom halves, such as softirq, tasklets, and work queues, by using an interrupt handling hook provided by \ostd. \ostd enforces ``atomic mode'' to prevent client-provided callbacks from sleeping in interrupt context. \aster also maintains system time, monotonic time, and wall clocks by registering timer interrupts and reading the timestamp counter (TSC) through \ostd.

\aster takes advantage of \ostd's safe policy injection feature. It incorporates a Linux-style task scheduler with multiple scheduling classes—including a real-time scheduler and a rudimentary CFS~\cite{CFSSchedulerLinux}. \aster provides an efficient and scalable buddy system frame allocator, with per-CPU caching. \aster also features a slab allocator following the original design~\cite{slabSingle}. Our primary focus has been establishing the injection mechanism rather than extensively refining each scheduling or allocation policy; we plan to continue improving these components for greater maturity.

\bheading{Performance optimization.}
Many performance bottlenecks in \aster and \ostd have been identified and addressed. Due to space constraints, we cannot cover every optimization in detail. In many instances, we have adapted strategies from Linux's optimized C implementations to align with the framekernel philosophy and the constraints of safe Rust.

One optimization in \aster is a pooling mechanism for DMA-able memory regions (akin to persistent mapping~\cite{protection-iommu}), which may be requested frequently by device drivers. This approach minimizes the need for DMA mapping setup, requiring it only during initialization, thereby preserving IOTLB entries and enhancing the hit rates for IOMMU. In contrast, a dynamic mechanism would necessitate frequent unmaps, leading to IOTLB invalidation and a subsequent performance decline.

Thus far, our optimization efforts have been concentrated on single-core systems. Ongoing work is focused on improving SMP scalability through the use of SMP-friendly locks (e.g., RCU~\cite{rcu}, MCS locks~\cite{mcs}) and the application of more fine-grained locking mechanisms. In this paper, our evaluation compares \aster directly with Linux in a single-core environment, with a comprehensive multi-core analysis planned for future work.

%% file: 6-eval.tex
\section{Evaluation}
\label{sec:eval}

\subsection{Performance Evaluation}

We evaluate the performance of \aster. \minor{The experiments are performed on a machine with an Intel i7-10700 processor, 32GB of memory, and an Intel SSDPEKNW512GB solid state drive. The system software on the machine includes Ubuntu 22.04 (Linux version 6.8.0) and QEMU 9.1.0.}

For comparison with \aster, we use Linux kernel version 5.15 as the baseline. Some Linux features that are missing in \aster, including CPU mitigations \major{(\texttt{mitigations=off})} and huge pages \major{(\texttt{hugepages=0})}, are disabled to ensure fairness. \major{Without disabling CPU mitigations, Linux's performance would be affected considerably.}

Both the Linux and \aster kernels are tested within QEMU virtual machines (VMs) configured as follows: SMP set to 1, machine type set to q35 with the kernel IRQ chip in split mode, CPU specified as Icelake-Server, PCID disabled, and x2APIC and KVM enabled. Each VM is attached with a virtio-blk device (having a single queue with a length of 64) and a virtio-net device (a tap device with vhost support enabled), which will be used for block and network I/O benchmarks, respectively.

\subsubsection{Micro-benchmarks}

We run LMbench~\cite{lmbench}, a series of system call-intensive microbenchmarks. Specifically, the benchmarks are classified into five categories: process-related (Proc), memory-related (Mem), Inter-process communication-related (IPC), filesystem-related (FS), and network-related benchmarks (Net). The last column (Norm) of Table~\ref{tab:basic-lmbench} shows the normalized performance---for throughput it shows results of Asterinas/Linux; for latency it shows results of Linux/Asterinas---and hence higher results suggest better performance of \aster. \aster is evaluated with IOMMU enabled and all reported results are averaged over \major{ten} runs of the benchmarks. 

As shown in Table~\ref{tab:basic-lmbench}, the geometric mean of the normalized performance is \minor{1.08}, which means \aster slightly outperforms Linux in more benchmarks. This result suggests that \aster is comparable to Linux in performance, but does not necessarily mean \aster is better optimized. As a newly developed Rust kernel, \aster misses some of the advanced features and configurations in Linux. For example, in the network TCP tests, \aster uses the smoltcp~\cite{smoltcp} library, which lacks congestion control, enabling TCP to operate at full speed and resulting in faster performance compared to Linux.
In the FS-related open and stat tests, Linux employs RCU-walk to achieve faster filename lookup, an optimization not implemented in \aster.

\begin{table}[!t]
    \centering
    \scriptsize
    \caption{System call-intensive LMbench microbenchmarks.}
    \begin{threeparttable}
    \begin{tabular}{l|l>{\hspace{-2mm}}cccc}
    \hline
    \rule{0pt}{8pt}
         & \textbf{Command} & \textbf{Unit} & \textbf{Linux} & \textbf{Asterinas} & \textbf{Norm.}\tnote{1} \\[2pt]
        \hline
            \multirow{5}{1.7em}{\scriptsize Proc} 
         & \texttt{lat\_syscall null}           & $\mu$s & 0.050 	 & 0.066 $\pm$ 0.001& 0.76 \\
         & \texttt{lat\_ctx 18}                 & $\mu$s & 0.826 	 & 0.829 $\pm$ 0.019& 1.00 \\
         & \texttt{lat\_proc fork}              & $\mu$s & 59.20  & 57.46 $\pm$ 0.721& 1.03 \\
         & \texttt{lat\_proc exec}              & $\mu$s & 204.8 & 174.4 $\pm$ 2.156& 1.17 \\
         & \texttt{lat\_proc shell}             & $\mu$s & 319.3 & 294.3 $\pm$ 1.915& 1.08 \\
         \hline
         
            \multirow{3}{1.7em}{Mem} 
         & \texttt{lat\_pagefault}              & $\mu$s & 0.109 	 & 0.100 $\pm$ 0.002& 1.09 \\
         & \texttt{lat\_mmap 4m}                & $\mu$s & 19.4  & 16.80 $\pm$ 0.422& 1.15 \\
         & \texttt{bw\_mmap\tnote{2} 256m}    & MB/s   & 15405  & 13197 $\pm$ 186.7& 0.86 \\
         \hline

            \multirow{5}{1.7em}{IPC} 
         & \texttt{lat\_pipe}                   & $\mu$s & 1.826 	 & 1.881 $\pm$ 0.009& 0.97 \\
         & \texttt{bw\_pipe}                    & MB/s   & 11133  & 14664 $\pm$ 1073& 1.32 \\
         & \texttt{lat\_fifo}                   & $\mu$s & 1.825 	 & 1.938 $\pm$ 0.008& 0.94 \\
         & \texttt{lat\_unix}                   & $\mu$s & 2.677 	 & 2.493 $\pm$ 0.023& 1.07 \\
         & \texttt{bw\_unix}                    & MB/s   & 7875   & 14183 $\pm$ 598.2& 1.80 \\
        \hline
         
            \multirow{10}{1.7em}{FS} 
         & \texttt{lat\_syscall open}           & $\mu$s & 0.611 	 & 0.740 $\pm$ 0.020& 0.83 \\
         & \texttt{lat\_syscall read}           & $\mu$s & 0.081 	 & 0.088 $\pm$ 0.002& 0.92 \\
         & \texttt{lat\_syscall write}          & $\mu$s & 0.065 	 & 0.080 $\pm$ 0.003& 0.81 \\
         & \texttt{lat\_syscall stat}           & $\mu$s & 0.299 	 & 0.400 $\pm$ 0.009& 0.75 \\
         & \texttt{lat\_syscall fstat}          & $\mu$s & 0.263 	 & 0.231 $\pm$ 0.004& 1.14 \\
         & \texttt{bw\_file\_rd\tnote{3} 512m}  & MB/s   & 10238 	 & 9198 $\pm$ 44.25& 0.90 \\
         & \texttt{lmdd(Ramfs->Ramfs)}        & MB/s   & 3219 	 & 2973 $\pm$ 18.60& 0.92 \\
         & \texttt{lmdd(Ramfs->Ext2)}       & MB/s   & 2490 	 & 2612 $\pm$ 57.08& 1.05 \\
         & \texttt{lmdd(Ext2->Ramfs)}       & MB/s   & 3453 	 & 2962 $\pm$ 30.77& 0.86 \\
         & \texttt{lmdd(Ext2->Ext2)}      & MB/s   & 2017 	 & 2626 $\pm$ 68.94& 1.30 \\
        \hline

            \multirow{4}{1.7em}{Net: Loop-back} 
         & \texttt{lat\_udp}         & $\mu$s & 3.801 	 & 2.427 $\pm$ 0.009& 1.57 \\
         & \texttt{lat\_tcp}         & $\mu$s & 5.326 	 & 2.725 $\pm$ 0.016& 1.95 \\
         & \texttt{bw\_tcp 128}      & MB/s   & 280.0	 & 356.5 $\pm$ 78.94& 1.27 \\
         & \texttt{bw\_tcp 64k}      & MB/s   & 6216 	 & 7647 $\pm$ 321.5& 1.23 \\
         
         \hline
         \multirow{4}{1.7em}{Net: VirtIO} 
         & \texttt{lat\_udp}       & $\mu$s & 15.03 	 & 11.49 $\pm$ 0.139& 1.31 \\
         & \texttt{lat\_tcp}       & $\mu$s & 16.75 	 & 12.94 $\pm$ 0.135& 1.29 \\
         & \texttt{bw\_tcp 128}    & MB/s   & 328.7 	 & 333.2 $\pm$ 5.141& 1.01 \\
         & \texttt{bw\_tcp 64k}    & MB/s   & 1151 	 & 1116 $\pm$ 15.05& 0.97 \\
        \hline
        \rule{0pt}{7pt}
        \textbf{Mean} &       & & 	 &   & \textbf{1.08} \\[1pt]
        \hline
        
    \end{tabular}
    \scriptsize
    \begin{tablenotes}[flushleft]
       \item[1] Normalized performance. For throughput, use Asterinas/Linux; for latency, use Linux/ Asterinas. The higher the better.
       \item[2] With mmap\_only.
       \item[3] With io\_only.
    \end{tablenotes}
    \end{threeparttable}
    \label{tab:basic-lmbench}
    \vspace{-0.6cm}
\end{table}

\subsubsection{Macro-benchmarks}

\pgfplotstableread[row sep=\\,col sep=&]{
file-size &  Linux     & Linux-err & Aster & Aster-err & Aster-no-IOMMU & Aster-no-iommu-err\\
4KB     & 19227.13  & 290.18 & 22912.4  & 232.72 & 22934.05  & \\
8KB     & 18105.02  & 328.08 & 20869.93 	& 207.34 & 20764.55 & \\
16KB    & 15716.4  & 293.10  & 18799.28 	& 104.91 & 18716.06 & \\
32KB    & 11882.7	& 169.04 & 15498.57 & 140.46 & 15519.04 & \\
64KB    & 9053.73	& 134.59 & 9233.79	& 132.39 & 9264.43 & \\
}\nginxdata

\pgfplotstableread[row sep=\\,col sep=&]{
operation & Linux & Aster & Aster-no-IOMMU \\
SET         & 153390.59 	& 211647.77 	& 210302.41 \\
GET         & 155994.34 	& 218670.04 	& 219300.26 \\
LPUSH       & 149886.75 	& 211691.68 	& 211960.12 \\
LPOP        & 148347.63 	& 209365.11 	& 209308.70 \\
LRANGE600   & 23876.49	    & 23649.05	    & 23674.88  \\
}\redisdata

\pgfplotstableread[row sep=\\,col sep=&]{
operation   & Linux     & Linux-err & Aster & Aster-err & Aster-no-IOMMU & Aster-no-iommu-err\\
120         & 0.88 	&      & 1.00 	&      & 1.00 & \\
200         & 1.59 	&      & 2.21 	&      & 2.07 &\\
230         & 1.81 	&      & 2.11 	&      & 2.08 &\\
400         & 1.44 	&      & 1.57 	&      & 1.58 &\\
410         & 2.25 	&      & 3.06 	&      & 3.05 &\\
}\sqliteexttwodata

\definecolor{Aster-color}{HTML}{b4e9ad}
\definecolor{Aster-no-iommu-color}{HTML}{fad7a0}
\definecolor{Graph-description-color}{HTML}{808080}
\definecolor{Linux-color}{HTML}{808080}
\begin{figure*}
    \centering
    \subfigbottomskip=2pt 
    \subfigcapskip=-5pt 

    \subfigure[Nginx]{
        \label{fig:application-bench:ngnix}
        \begin{tikzpicture}[scale=0.83]
        \begin{axis}[
        compat=newest,
            ybar,
            bar width=3.5pt,
            width=.42\textwidth,
            height=.22\textwidth,
            ybar legend,
            legend cell align=center,
            legend style={
                cells={anchor=center},
                at={(0.5,1.0)},    
                anchor=north,
                legend columns=-1,                
                inner ysep=.5pt, 
                },
            legend image code/.code={%
                \draw[#1] (0.0cm,-0.064cm) rectangle (0.07cm,0.1cm); 
            },        
            symbolic x coords={4KB,8KB,16KB,32KB,64KB},
            xtick=data,
            ymin=0,
            ymax=35000,
            xlabel={File size},
            ylabel={Throughput (rps)},
            enlarge x limits=0.17, 
            clip=false,
            xtick style={draw=none}, 
            error bars/y dir=both, 
            error bars/y explicit,
            error bars/error bar style={red, thin}, 
        ]
        \addplot 
        [fill=Linux-color,error bars/.cd,y dir=both,y explicit] 
        table[x=file-size,y=Linux]{\nginxdata};
        \addplot 
        [fill=Aster-color,error bars/.cd,y dir=both,y explicit]
        table[x=file-size,y=Aster]{\nginxdata};
        \addplot 
        [fill=Aster-no-iommu-color]
        table[x=file-size,y=Aster-no-IOMMU]{\nginxdata};
        \scriptsize
        \legend{Linux, Asterinas, Asterinas-no-IOMMU}
        \end{axis}
        \end{tikzpicture}
        }
    \subfigure[Redis]{
        \label{fig:application-bench:redis}
        \begin{tikzpicture}[scale=0.83]
        \begin{axis}[
        compat=newest,
            ybar,
            bar width=3.5pt,
            width=.42\textwidth,
            height=.22\textwidth,
            legend cell align=center,
            legend style={
                cells={anchor=center},
                at={(0.5,1.0)},    
                anchor=north,
                legend columns=-1,
                inner ysep=.5pt, 
                },
            legend image code/.code={%
                \draw[#1] (0.0cm,-0.064cm) rectangle (0.07cm,0.1cm); 
            },        
            symbolic x coords={
            SET,GET,LPUSH,LPOP,
            LRANGE600
            },
            xtick=data,
            ymin=0,
            ymax=280000,
            xlabel={Operation},
            ylabel={Throughput (rps)},
            enlarge x limits=0.14, 
            clip=false,
            xtick style={draw=none}, 
        ]
        \addplot 
        [fill=Linux-color] 
        table[x=operation,y=Linux]{\redisdata};
        \addplot 
        [fill=Aster-color]
        table[x=operation,y=Aster]{\redisdata};
        \addplot 
        [fill=Aster-no-iommu-color]
        table[x=operation,y=Aster-no-IOMMU]{\redisdata};
        \scriptsize 
        \legend{Linux, Asterinas, Asterinas-no-IOMMU}
        \end{axis}
        \end{tikzpicture}
        }
    \subfigure[\footnotesize SQLite]{
        \label{fig:application-bench:sqlite}
        \begin{tikzpicture}[scale=0.83]
        \begin{axis}[
        compat=newest,
            ybar,
            bar width=3.5pt,
            width=.42\textwidth,
            height=.22\textwidth,
            legend cell align=center,
            legend style={
                cells={anchor=center},
                at={(0.5,1.0)},    
                anchor=north,
                legend columns=-1,
                inner ysep=.5pt, 
                },
            legend image code/.code={%
                \draw[#1] (0.0cm,-0.064cm) rectangle (0.07cm,0.1cm); 
            },        
            symbolic x coords={120,200,230,400,410},
            xtick=data,
            ymin=0,
            ymax=4.2,
            xlabel={Test Number},
            ylabel={Latency (s)},
            clip=false,
            xtick style={draw=none}, 
            enlarge x limits=0.17, 
        ]
        \addplot 
        [fill=Linux-color]
        table[x=operation,y=Linux]{\sqliteexttwodata};
        \addplot 
        [fill=Aster-color]
        table[x=operation,y=Aster]{\sqliteexttwodata};
        \addplot 
        [fill=Aster-no-iommu-color]
        table[x=operation,y=Aster-no-IOMMU]{\sqliteexttwodata};
        \scriptsize 
        \legend{Linux, Asterinas, Asterinas-no-IOMMU}
        \end{axis}
        \end{tikzpicture}
        }
        \vspace{-0.2cm}
        \caption{I/O-intensive application benchmarks with Ngnix, Redis, and SQLite.}
        \label{fig:application-bench}
        \vspace{-0.35cm}
\end{figure*}
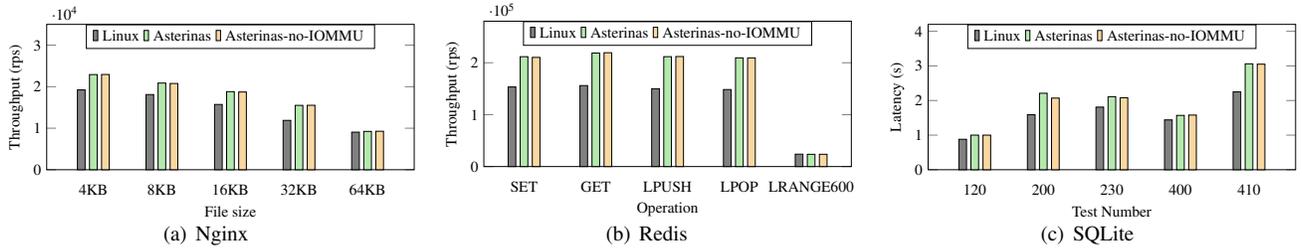

We evaluate \aster's performance with popular I/O-intensive applications, including Nginx (1.26.2),  Redis (7.0.15), and SQLite (3.46.1). The results in Figure \ref{fig:application-bench} report \aster's performance with IOMMU enabled (default) and disabled (for comparison). 

\begin{packeditemize}
\item \textbf{Nginx.} We evaluate the throughput of Nginx using ApacheBench~\cite{apache-bench} with a concurrency level of 32 and a total of 200,000 requests. The results shown in Figure~\ref{fig:application-bench:ngnix} indicate that the throughput of Nginx is higher on \aster
\minor{(22,912 rps) than Linux (19,227 rps) when the requested file size is 4096 bytes (19\% higher).} \major{When the file size is 64 KiB, \aster's throughput drops to 9,234 RPS, which is close to Linux’s result. This drop may happen because \aster’s \texttt{sendfile} implementation is less optimized—it requires an extra copy to an intermediate buffer. For small files (4 KiB), \aster performs better than Linux because it lacks congestion control (using the smoltcp crate). However, as the file size increases, the overhead of redundant copying outweighs this advantage, causing \aster’s performance to fall behind Linux.}
The geometric mean of all the normalized performances (\aster / Linux) is \minor{1.17}.

\item \textbf{Redis.} We use the official Redis benchmark tool~\cite{redis-benchmark}, which measures performance by executing a series of Redis commands and recording the throughput. Figure~\ref{fig:application-bench:redis} shows a subset of representative commands, and the complete results can be found in Appendix~\ref{appendix:redis-full-results}. The results indicate that \aster outperforms Linux. \minor{For instance, the throughput for GET operation on \aster is 218,670 rps, 40.2\% higher compared to 155,994 rps on Linux.} This result aligns with our LMbench findings, where \aster outperforms Linux for smaller packet sizes, as the Redis benchmark involves small message sizes in its operations. The geometric mean of all the normalized performance results is \minor{1.31}. 

\item \textbf{SQLite.} We utilize SQLite's speedtest1~\cite{sqlite-speedtest1} as the benchmarking tool. In the test, we set the base size to 1000 to ensure data is written to the device (default is 100) and run the tests on an Ext2 mount over a virtio-blk device. Due to space limits, we only show representative results in Figure~\ref{fig:application-bench:sqlite}, whose test numbers are 120 (500000 unordered INSERTS with one index/PK), 200 (VACUUM), 230 (100000 UPDATES, numeric BETWEEN, indexed), 400 (700000 REPLACE ops on an IPK), and 410 (700000 SELECTS on an IPK). The complete results are shown in Appendix~\ref{appendix:sqlite-full-results}. The geometric mean of all the normalized performances (Linux / \aster) is \minor{0.85}.

In all these tests, \aster performs worse than Linux. The Vacuum test (200) shows the worst case, reaching only \minor{72\%} of Linux's performance with IOMMU enabled. Once the IOMMU is disabled, the ratio will become \minor{77\%}. To understand the root causes of Vacuum test, we used Strace~\cite{strace} for performance diagnosis. The analysis revealed that Vacuum frequently makes \texttt{pwrite64} calls for writing 4-byte data. When disabling the SQLite journal, \texttt{pwrite64} with small data writes disappears, and \aster' performance increases to \minor{80\% and 83\%} of Linux's performance with IOMMU enabled and disabled, respectively. This analysis shows that \aster could benefit from optimization for both \texttt{pwrite64} operations with small data writes and IOMMU.

\end{packeditemize}

\pgfplotstableread[row sep=\\,col sep=&]{
        Label               & base     & overhead & total \\
        Bytes Read (4KB)    & 97.79 &  2.21  & 124.569 \\
        Bytes Write (4KB)   & 99.41 &  0.59  & 238.788 \\
        Device Read (32bit)   & 98.45 &  1.55  & 10988.49 \\
        Device Write (32bit)  & 98.44 &  1.56  & 10665.73 \\
        Create Stack          & 96.08 &  3.92 & 2851.49 \\
        Yield Task     & 99.5 & 0.5 & 167.11 \\
        Allocate Frame     & 90 & 10 & 775 \\
        Allocate Object      & 90 & 10 & 875 \\
}\safetyoverheaddata

\begin{table}[t]
    \centering
    \footnotesize
    \caption{Overhead due to \ostd's safety mechanism.}
    \begin{tabular}{l@{\hspace{4pt}}l@{\hspace{4pt}}c}
        \toprule
         \multirow{2}[3]{*}{\textbf{Operations}} &
         \multirow{2}[3]{*}{\makecell{ \textbf{Sources of} \\ \textbf{Safety Overheads}}} &
         \textbf{CPU Cycles} \\ 
         \cmidrule(lr){3-3} & & \textbf{Overhead / Total} \\
         \midrule
        Segment::read\_bytes (4KB)     & Boundary check   & 3/125(2.4\%)      \\
        Segment::write\_bytes (4KB)    & Boundary check   & 2/239(0.8\%)      \\
        IoMem::read\_once (4 bytes) & Boundary check   & 170/10988(1.5\%)    \\
        IoMem::write\_once (4 bytes) & Boundary check   & 166/10666(1.6\%)    \\
        KernelStack::new         & Guard page creation    & 25/2950(0.8\%)     \\
        Task::yield\_now           & Running flag check      & 1/167(0.6\%)      \\
        FrameAlloc::alloc (1 frame)       & Ownership check        & 12/180(6.7\%)     \\
        Box::new (48 bytes)      & Ownership check        & 1/148(0.7\%) \\
        \hline
    \end{tabular}
    \footnotesize
    \label{tab:safety-overhead}
    \vspace{-0.4cm}
\end{table}

\subsubsection{Overhead due to Safety Checks}
\label{subsec:eval:safety}
The overhead introduced by the \ostd's safety mechanisms does not prevent \aster's performance from being comparable to a monolithic kernel. \aster has integrated several safety mechanisms into its \ostd to ensure system safety. To evaluate the cost of safety mechanisms, we directly invoke the interfaces corresponding to these safety mechanisms at the API level, measuring the CPU cycles required with and without the safety checks. The results are presented in Table~\ref{tab:safety-overhead}. The "Allocate frame" test, which is related to the Frame allocator injection, exhibits the highest overhead at 6.7\%, while the overhead of all other tests remains below 3\%.

\subsubsection{IOMMU Optimization}

One optimization of IOMMU using the pooling mechanism leads to a significant performance gain. As shown in Figure~\ref{fig:application-bench}, the results indicate that \aster does not introduce significant overhead for Nginx, Redis, and SQLite (in most cases). \minor{Both Nginx and Redis show almost no overhead, with the largest overhead observed in the SPOP test of Redis at 2.3\%.}

\begin{figure}
    \centering
    \definecolor{without-tlb-flush-color}{HTML}{F9E79F}
    \newenvironment{customlegend}[1][]{%
        \begingroup
        \csname pgfplots@init@cleared@structures\endcsname
        \pgfplotsset{#1}%
    }{%
        \csname pgfplots@createlegend\endcsname
        \endgroup
    }%
    \def\addlegendimageonly{\csname pgfplots@addlegendimage\endcsname}

    \hspace*{.07\textwidth}
    \begin{tikzpicture}
    \scriptsize
    \centering
        \begin{customlegend}[
        legend columns=-1,
        legend style={
            align=center,
            draw=none,
            column sep=0.1cm,
        },
        legend entries={
            {Dynamic mapping},
            {Pooling - static mapping},
        }]
            \addlegendimageonly{
                legend image code/.code={
                    \draw[black, fill=Aster-no-iommu-color] (0.0cm,-0.064cm) rectangle (0.08cm,0.13cm);
                }
            } 
            \addlegendimageonly{
                legend image code/.code={
                    \draw[black, fill=Aster-color] (0.0cm,-0.064cm) rectangle (0.08cm,0.13cm);
                }
            }
        \end{customlegend}
    \end{tikzpicture}
    \vspace{-0.35cm}

    \subfigure[Block device bandwidth]{
        \centering
        \begin{tikzpicture}
        \begin{axis}[
            compat=newest,
            ybar,
            bar width=5.5pt,
            width=.25\textwidth,
            height=.18\textwidth,
            legend style={
                at={(0.5,1)},
                anchor=north,
                legend columns=-1,
                },
            xtick=data,
            symbolic x coords={Seq Read,Seq Write},
            ymin=0,
            xlabel={Operation},
            ylabel={MB/s},
            enlarge x limits=0.6,
            clip=false,
            xtick style={draw=none}, 
        ]
        \addplot 
        [fill=Aster-no-iommu-color]
        table[x=operation,y=dynamic]{\iommublock};
        \addplot 
        [fill=Aster-color]
        table[x=operation,y=static]{\iommublock};
        \scriptsize 
        \end{axis}
        \end{tikzpicture}
    }
    \subfigure[Network device bandwidth]{
        
        \begin{tikzpicture}
        \begin{axis}[
            compat=newest,
            ybar,
            bar width=5.5pt,
            width=.25\textwidth,
            height=.18\textwidth,
            legend style={
                at={(0.5,1)},
                anchor=north,
                legend columns=-1,
                },
            xtick=data,
            symbolic x coords={128,65536},
            ymin=0,
            xlabel={Message Size},
            ylabel={MB/s},
            enlarge x limits=0.6,
            clip=false,
            xtick style={draw=none}, 
        ]
        \addplot 
        [fill=Aster-no-iommu-color]
        table[x=operation,y=dynamic]{\iommunet};
        \addplot 
        [fill=Aster-color]
        table[x=operation,y=static]{\iommunet};
        \scriptsize 
        \end{axis}
        \end{tikzpicture}
    }
    \vspace{-0.2cm}
    \caption{IOMMU overhead with different mechanism}
    \label{fig:iommu-overhead}
    \vspace{-0.2cm}

\end{figure}
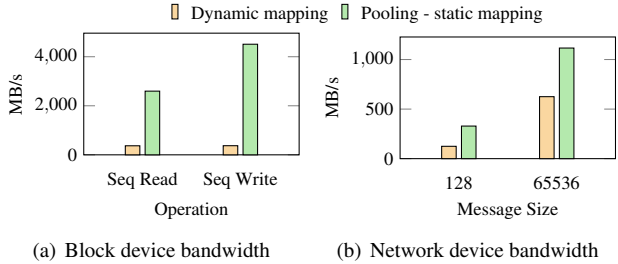

However, SQLite exhibits higher overhead in some cases (\minor{8.3\%}), primarily due to the incomplete pooling mechanism in the block device driver compared to the network device driver.
Figure~\ref{fig:iommu-overhead} further highlights the significance of pooling. We evaluate the bandwidth of block devices using FIO~\cite{fio} and network devices using bw\_tcp. Upon switching from polling to a dynamic mechanism, both network and block devices experienced considerable performance degradation.

\pgfplotstableread[row sep=\\,col sep=&]{
operation & static & dynamic \\
Seq Read     & 2597 & 369\\
Seq Write    & 4510 & 372 \\
}\iommublock

\pgfplotstableread[row sep=\\,col sep=&]{
operation & static  & dynamic\\
128   & 328.7 & 124.58  \\
65536   & 1116 & 625.33 \\
}\iommunet

\subsection{TCB Evaluation}

In this section, we evaluate the current TCB size of \aster and estimate its growth as the codebase evolves.

\subsubsection{TCB Comparison}
To assess the efficacy of \aster's implementation of the minimality principle, we compare the TCB size of \aster with RedLeaf~\cite{RedLeaf}, Theseus~\cite{Theseus}, and Tock~\cite{Tock}. 

\bheading{Methodology.}
We employ a crate-level classification approach to evaluate the run-time TCBs of Rust-based OSes. In Rust, crates serve as the primary units for organizing and distributing code, with no additional isolation mechanisms within a crate. Consequently, a crate's interface naturally forms a trust boundary. So we consider a crate either belongs to the TCB, implying it must be trusted for security, or is excluded from the TCB, indicating it is secured by the Rust compiler. We adhere to the following rules to determine whether a crate is within the run-time TCB or not:

\begin{packeditemize}
\item \textbf{Rule 1}: The Rust toolchain is trusted. Consequently, crates provided by the Rust toolchain itself, like \texttt{alloc} and \texttt{core}, are \emph{not} considered part of the run-time TCB.

\item \textbf{Rule 2}: Crates containing \texttt{unsafe} code may potentially introduce soundness bugs, and as a result, they are considered part of the TCB.

\item \textbf{Rule 3}: Crates that are dependencies of TCB crates should also be part of the TCB. This is because even if a crate doesn't use \texttt{unsafe} at all and thus doesn't introduce soundness issues on its own, the correctness of its APIs can still impact the soundness of the TCB crate that relies on it.
\end{packeditemize}

Since crates vary in size, evaluating the TCB size by counting the number of crates would be too simplistic and inaccurate. Therefore, a more refined metric is needed. However, directly comparing lines of code across crates is not ideal, as not all code within a crate is necessarily utilized during runtime. This is especially true for third-party dependencies, where often only a small fraction of the code is actually used at runtime. 

To address this, we introduce a metric called \textbf{Linked Code Size (LCS)}, which measures the number of lines of code that are ultimately compiled and linked during the OS build. 

We leverage the LLVM toolchain~\cite{llvmtools2025} to estimate the LCS for each OS. Specifically, we measure the number of source code lines that have corresponding statements in the generated LLVM IRs after link-time optimization. The content of the LLVM IR is closer to that of a binary file, primarily retaining the IR corresponding to executable statements within functions. It does not trace back to statements used for imports or struct definitions, which are not directly executable during runtime. This focus on executable IR provides a more accurate measure of the code that is relevant to the TCB during runtime operations.

\bheading{Results.}
As illustrated in Table~\ref{table:TCB}, the relative TCB size of \aster is a mere 14.0\%, which is notably lower than that of other Rust-based OSes. While some OSes, such as RedLeaf, enforce restrictions that permit only safe Rust in specific crates, they still depend on numerous self-maintained crates that utilize \texttt{unsafe}, leading to a substantial portion of their code being part of the TCB (i.e., 66.1\%). 

Tock OS, being an embedded OS, is relatively lightweight, and thus its TCB size is slightly better than the other two. However, its TCB size (i.e., 43.8\%) is still more than double of \aster. These results clearly underscore the effectiveness of the minimality principle that we adhere to.

\begin{table}[t]
 \centering
\caption{Comparison of TCB size}
\footnotesize
 \begin{threeparttable}
\begin{tabular}{@{}ccccc@{}}
\toprule
 & RedLeaf & Theseus & Tock\tnote{1} & \textbf{Asterinas} \\ 
\midrule
Total (LoC) & 25992 & 70468 & 6628 & \textbf{75285} \\
TCB (LoC) & 17182 & 43978 & 2903 & \textbf{10571} \\
Relative TCB & 66.1\% & 62.4\% & 43.8\% & \textbf{14.0\%} \\ \bottomrule
\end{tabular}%
\label{table:TCB}
\scriptsize
    \begin{tablenotes}[flushleft]
       \item[1] Tock implementation of board \texttt{nrf52840dk}.
    \end{tablenotes}
\end{threeparttable}
\vspace{-0.3cm}
\end{table}

\subsubsection{TCB Evolution}

\begin{figure}[t]
  \centering
  \includegraphics[width=\linewidth]{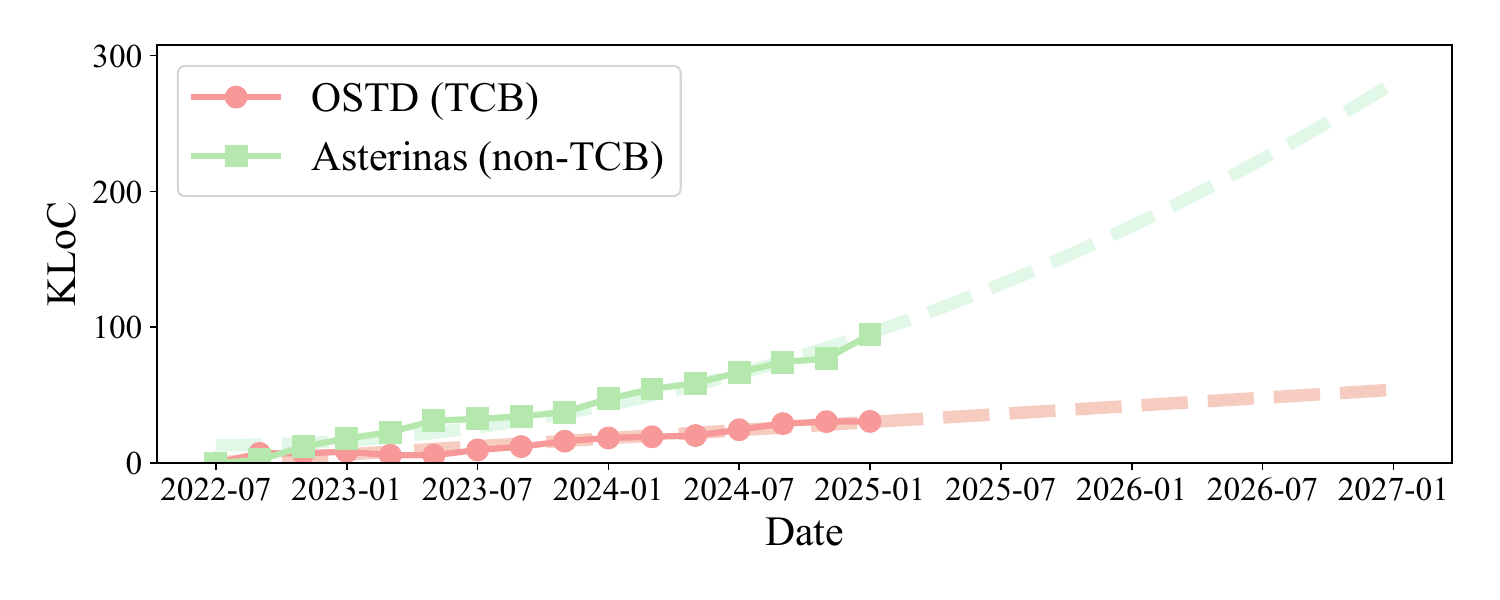}
  \vspace{-1.0cm}
  \caption{
  The codebase of \aster (non-TCB) has grown faster than \ostd (TCB) and is expected to remain so.}
  \label{fig:TCB}
  \vspace{-0.2cm}
\end{figure}

\aster has been under development for three years. We analyze and visualize the growth of its codebase (in KLoC). As shown in Figure~\ref{fig:TCB}, \aster (non-TCB) has experienced significant growth, while \ostd (TCB) has maintained a slower, steadier expansion. Curve fitting indicates that the TCB portion of the kernel will continue to grow in a controlled manner. This aligns with prior research on Linux codebase development~\cite{understandingTCBevolution, GrowthLinux}, which suggests that while overall codebases grow super-linearly with increasing complexity, core kernel functionality—excluding drivers—grows sub-linearly~\cite{FreeBSDandLinux}. 

\subsection{Safety Evaluation}
\begin{figure}[t]
  \centering
  \hspace{-0.2cm}
  \includegraphics[width=0.9\linewidth]{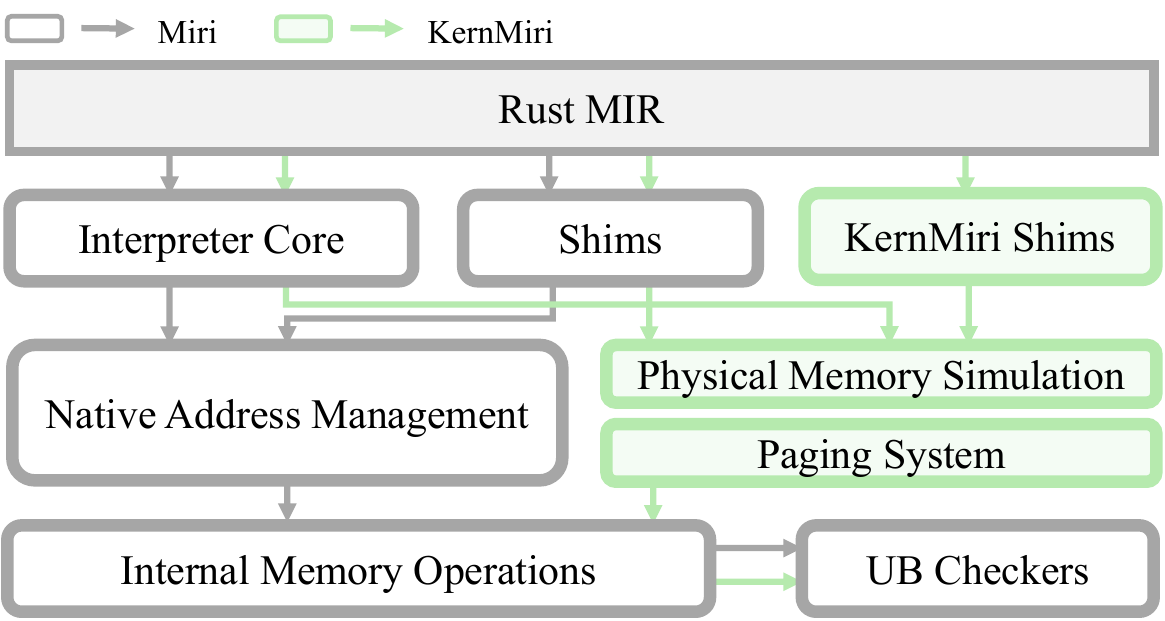}
  \vspace{-0.2cm}
  \caption{
  \kernmiri, a UB detection tool for Rust OSes
  }
  \label{fig:kern_miri}
  \vspace{-0.2cm}
\end{figure}

To evaluate the soundness of \ostd, we develop a testing tool named \textit{\kernmiri}. This tool extends the official Rust UB detection tool, Miri~\cite{miri}, to support core OS concepts such as physical memory, page tables, etc. The overall architecture of \kernmiri is depicted in Figure~\ref{fig:kern_miri}. It retains Miri's original UB detection logic (represented by grey boxes and lines) while introducing an additional 1,200 LoC to implement new components (indicated by green boxes and lines).

As illustrated in Figure~\ref{fig:kern_miri}, 
\kernmiri enhances Miri by simulating physical memory and implementing a basic paging system. These improvements enable \kernmiri to accurately interpret OS operations that demand fine-grained memory management. Moreover, \kernmiri introduces additional shims to synchronize the OS state, which can be leveraged to direct \kernmiri to activate a page table at a specific address or to inform \kernmiri of changes in the state of a physical page, such as transitions between typed and untyped states.

\kernmiri interprets the execution of \ostd and all its test cases. Following the workflow of \ostd's unit testing, \kernmiri first interprets the initialization phase of \ostd to synchronize the system's initial memory state, and then executes each unit test of \ostd via interpretation. As such, the coverage of \kernmiri depends on the unit test cases that have been developed for \ostd. 

\subsubsection{Coverage of \kernmiri} 

We primarily use \kernmiri to test and validate the \texttt{mm} module of \ostd, which is closely related to memory safety. This module also contains the majority of \texttt{unsafe} code in \ostd.
Table~\ref{table:unit_test} summarizes the measured metrics for each submodule of \texttt{mm} module separately, including the coverage of lines of code (Column \texttt{Line}), coverage of unsafe blocks (Column \texttt{Unsafe}) and interpreted execution time compared to native execution (Column \texttt{Execution}). 

The results suggest that, with in total 134 unit tests, \kernmiri covers all instances of \texttt{unsafe} code and over 90\% on average of lines in all submodules. This confirms that the majority of the \ostd's memory operations is free of UBs. Furthermore, the few detected instances of UB have been addressed, with a detailed discussion provided in Section~\ref{sec:ubcase}.

In addition, as shown in the last column of Table~\ref{table:unit_test}, \kernmiri's interpreted execution takes approximately 25 times longer than normal execution. While significantly higher, this overhead is reasonable as a one-time cost for soundness evaluation. Thus, we conclude that \kernmiri is a practical and effective tool for validating the soundness of Rust-based operating systems.  

\newcolumntype{C}{>{$}c<{$}}

\begin{table}[t]
\centering
\caption{Coverage and efficiency of \kernmiri on \ostd.}
\scriptsize
\begin{tabular}{@{}l@{\hspace{5pt}}c@{\hspace{5pt}}cccc@{}}
\toprule
\multirow{2}{*}{Modules} & \multirow{2}{*}{\# Tests} & Lines & Unsafe & \multicolumn{2}{c}{Execution} \\
\cmidrule(lr){3-3}\cmidrule(lr){4-4}\cmidrule(lr){5-6}
 & & Covered / Total & Covered / Total & Native & KernMiri \\
 \midrule
 dma  & 12 & \makebox[5em][C]{385/443}\hfill(87\%) & \makebox[3em][C]{8/8}\hfill(100\%) & 0.25s & 1.22s \\
 frame & 28 & \makebox[5em][C]{634/649}\hfill(98\%) & \makebox[3em][C]{41/41}\hfill(100\%) & 0.21s & 3.14s \\
 heap & 6 & \makebox[5em][C]{278/319}\hfill(87\%) & \makebox[3em][C]{6/6}\hfill(100\%) & 0.01s & 0.31s \\
 kspace & 8 & \makebox[5em][C]{287/323}\hfill(89\%) & \makebox[3em][C]{9/9}\hfill(100\%) & 0.04s & 0.93s \\
 page\_table & 34 & \makebox[5em][C]{931/1032}\hfill(90\%) & \makebox[3em][C]{46/46}\hfill(100\%) & 1.23s & 34.83s \\
 io & 29 & \makebox[5em][C]{358/371}\hfill(97\%) & \makebox[3em][C]{23/23}\hfill(100\%) & 0.16s & 3.12s \\
 vm\_space & 17 & \makebox[5em][C]{662/672}\hfill(99\%) & \makebox[3em][C]{10/10}\hfill(100\%) & 0.28s & 6.95s \\
 \textbf{All} & \textbf{134} & \makebox[5em][C]{\textbf{3535/3809}}\hfill(93\%) & \makebox[3em][C]{\textbf{145/145}}\hfill(100\%) & \textbf{2.18s} & \textbf{50.50s} \\
 \bottomrule
 \vspace{-0.7cm}
\end{tabular}%

\label{table:unit_test}
\end{table}

\subsubsection{Case Studies}
\label{sec:ubcase}

\begin{figure}[t]
  \centering
  \subfigure[Data Race UB]{
    \includegraphics[width=\linewidth]{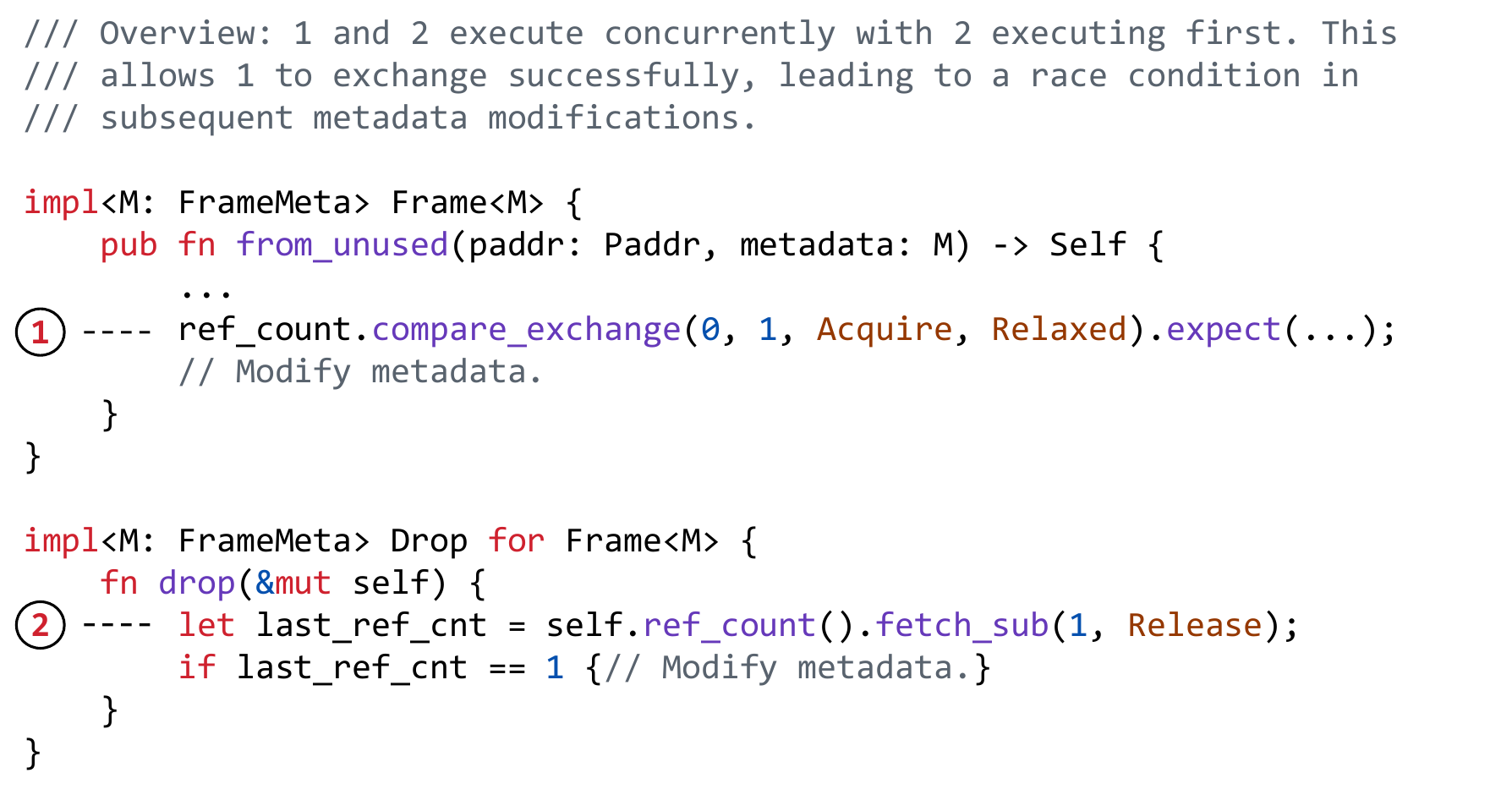}
    \label{fig:case_1}
  }
  \hfill
  \vspace{-0.2cm}
  \subfigure[Mutability UB]{
    \includegraphics[width=\linewidth]{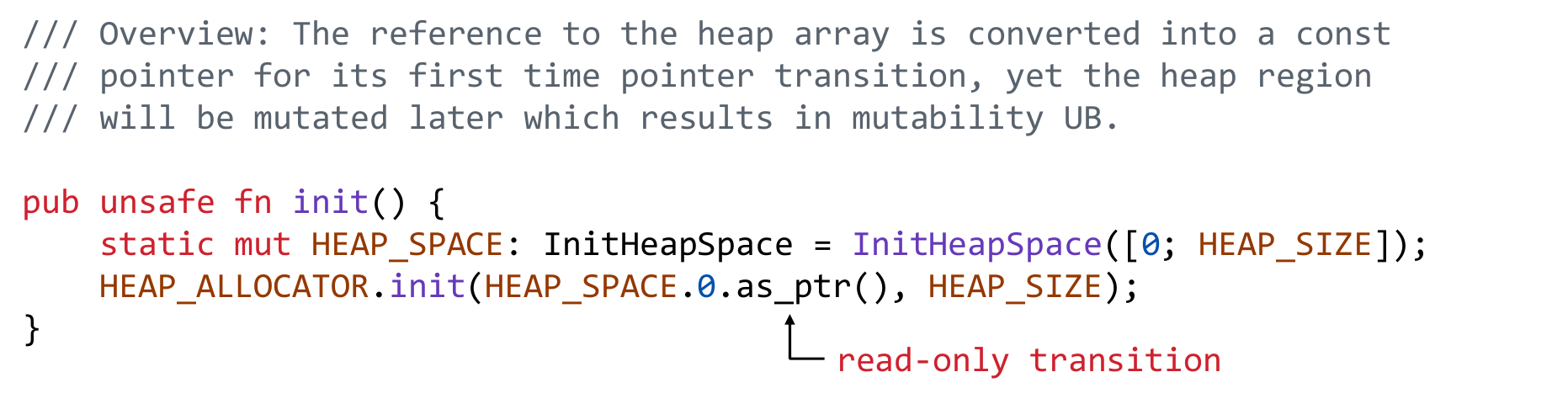}
    \label{fig:case_2}
  }
  \vspace{-0.2cm}
  \caption{UB cases detected by \kernmiri.}
  \vspace{-0.2cm}
  \label{fig:ub_cases}
\end{figure}

\kernmiri has helped us detect several UBs that we would not be able to catch using other methods, significantly improves the soundness of our implementation of \ostd. We outline two representative UB cases in \ostd detected by \kernmiri, as illustrated in Figure~\ref{fig:ub_cases}. 

The first case involves a data race UB (Figure~\ref{fig:case_1}). When creating a \texttt{Frame} using \texttt{from\_unused} (see~\ref{subsubsec:frame_allocator_injection}), if the \texttt{ref\_count} corresponding to the physical address has just been decremented by a \texttt{drop} operation that has not yet completed, these two operations may concurrently modify the metadata, leading to data race UB.

The second case pertains to mutability UB (Figure~\ref{fig:case_2}), where the code overlooks Rust's rules regarding the conversion of references to pointers. Specifically, when a reference is first converted into a pointer, the type of pointer determines the mutability constraints of the memory region. In this case a reference to the \texttt{HEAP\_SPACE} is incorrectly converted into an immutable pointer during initialization. This conversion conflicts with subsequent mutable operations, resulting in UB. 

%% file: 7-related.tex
\section{Related Work}

\textbf{Safe language-based OSes.} The development of operating systems using safe programming languages has been steadily increasing. Examples include Biscuit~\cite{Biscuit}, implemented in Go; JX~\cite{Jx}, developed in Java; MirageOS~\cite{MirageOS}, built with OCaml; Verve~\cite{Verve}, written in C\#; and Singularity~\cite{Singularity}, which utilizes Sing\# (an extension of C\#). Rust has also become a popular choice for system development, as seen in projects like Theseus~\cite{Theseus}, Tock~\cite{Tock}, RustyHermit~\cite{RustyHermit}, Redox~\cite{Redox}, and rCore~\cite{rCore}. \aster is also implemented in Rust but goes further by leveraging Rust's type and memory safety features to achieve a minimal and sound TCB.

\bheading{Intra-kernel privileged separation.} Various mechanisms have been proposed to achieve privilege separation within the kernel. PerspicuOS~\cite{nested-kernel} enforces separation using page table write protection and static analysis. RustyHermit-MPK~\cite{Isolation-Intel-MPK}, CubicleOS~\cite{CubicleOS}, and KDPM~\cite{KDPM} employ Intel MPK for isolation. Lightweight Virtualized Domains~\cite{lightweight-isolation} achieve separation through EPT and VMFUNC. RedLeaf~\cite{RedLeaf}, Tock~\cite{Tock}, and Theseus~\cite{Theseus} utilize Rust's safety features to create a new privilege level within the kernel. \aster achieves both soundness and minimality of intra-kernel privilege separation through a redefinition of the essential TCB functions. 

\noindent \textbf{Scheduler Injection.} ghOSt~\cite{ghOSt} modifies Linux to delegate scheduling to userspace, incurring non-trivial overhead, while Enoki~\cite{Enoki} adapts ghOSt's architecture but moves scheduler back to the kernel. In contrast, \aster offers (i) a smaller TCB—ghOSt relies on C-based run-queues, and Enoki can't catch all semantic bugs—and (ii) more flexible APIs, as Enoki schedulers must coordinate with Linux to avoid inconsistencies when managing run-queues, limiting independence.

%% file: 8-conclusion.tex
\section{Conclusions}

\minor{This paper presents \aster, a Linux ABI-compatible OS kernel based on our novel \emph{framekernel} architecture. By harnessing Rust's ownership and type-safety guarantees, a framekernel enforces \emph{intra-kernel privilege separation} to achieve a small TCB in terms of memory safety. Using the APIs of \ostd (TCB), \aster (non-TCB) is implemented entirely in safe Rust, supporting over 210 Linux system calls. Our comprehensive evaluation shows that \aster delivers performance on par with Linux, demonstrating that a fully-featured, general-purpose OS can be both memory-safe and highly efficient.}
\vspace{-4pt}

%% file: 9-acknowledgement.tex
\section{Acknowledgments}
\major{
The SUSTech authors are affiliated with the Research Institute of Trustworthy Autonomous
Systems and in part supported by National Key R\&D Program of China (No. 2023YFB4503902), CCF-AFSG RF20220012 and CCF-AFSG RF20230211.
The Peking University authors are in part supported by National Science Foundation of China (No. 62032001).
Asterinas has benefited from the collective expertise and support of numerous colleagues including
Tao Xie,
Bo An,
Lingxin Kong,
Yao Guo,
Leye Wang,
Huashan Yu,
Jie Zhang,
Wenfei Wu,
Hui Xu,
Zhaozhong Ni,
Wei Zhang,
Zhengyu He,
Tao Wei,
Lin Huang,
Chuan Song,
Yu Chen,
Weijie Liu,
Zhi Li,
and many more.
Special recognition is due to our open source contributors - 
Qing Li, Fabing Li, Shaowei Song, Qingsong Chen,
Siyuan Hui,
Zejun Zhao,
Qihang Xu,
Anmin Liu,
Wang Siyuan,
Wenqian Yan,
Zhenchen Wang,
Yingdi Shan,
Hu Kang,
Zhe Tang,
Ruize Tang,
and many more -
whose technical dedication and collaborative spirit transformed 
this ambitious vision into a reality.}

%% file: appendix.tex
\appendix

\section{Another Example for \ostd APIs}

\label{appendix:ostd_example2} 

\begin{figure}[H]
  \centering
  \includegraphics[width=\linewidth]{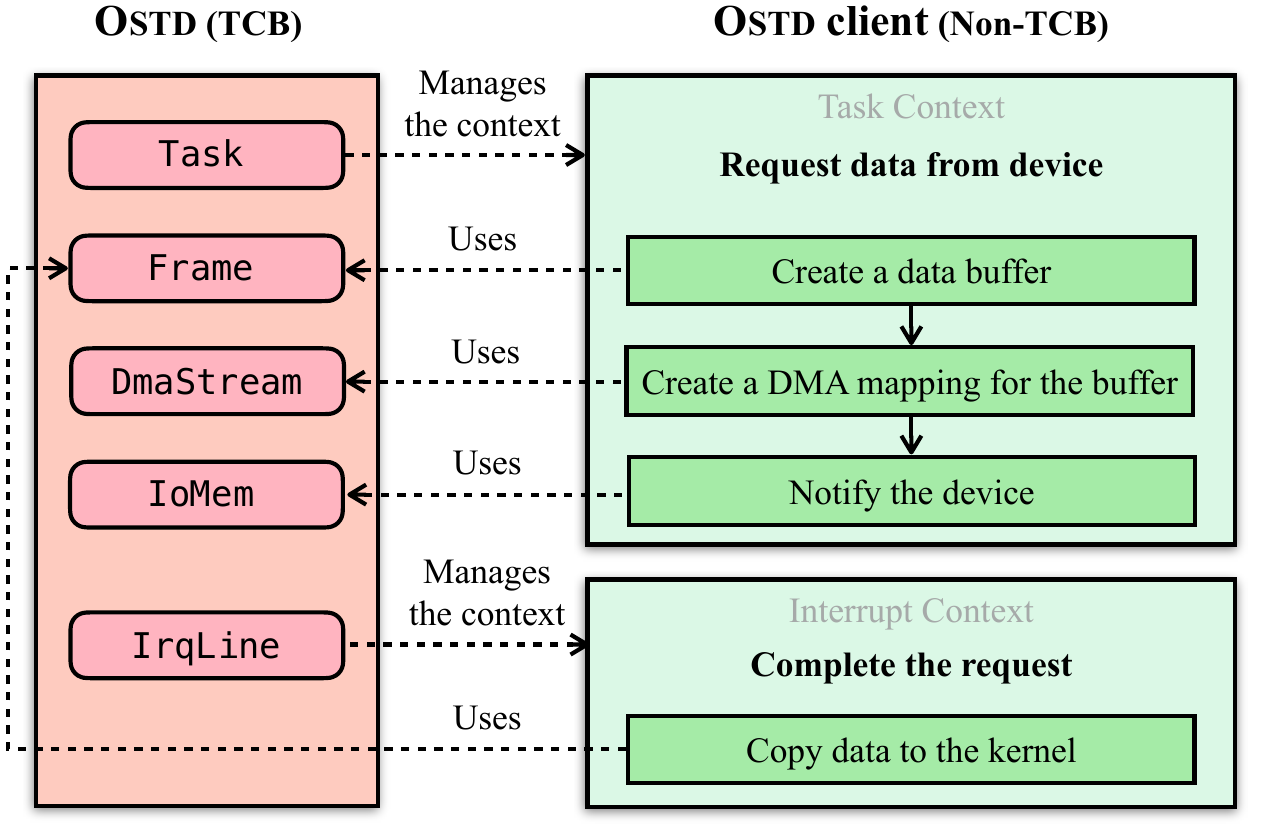}
  \vspace{-0.2cm}
  \caption{An example of \ostd API usage: request data from a device.}
  \label{fig:ostd_example2}
\end{figure}

\section{Complete results for Redis benchmark}
\label{appendix:redis-full-results}

Table~\ref{app:complete:redis} displays the complete results from testing Redis using the redis-benchmark. The left column lists all the operations conducted in the benchmark, while the three columns on the right present performance metrics for Linux, \aster, and \aster without IOMMU, respectively. The performance metrics are measured in requests per second (rps).

\begin{table}[H]
    \centering
    \footnotesize
    \begin{threeparttable}
    \begin{tabular}{l|ccc}
    \hline
    \textbf{Operation}& \textbf{Linux(rps)} & \textbf{Asterinas(rps)} & \makecell{\textbf{Asterinas}\\ \textbf{no IOMMU(rps)}}\\
        \hline
        PING\_INLINE  & 151022.44 	& 213341.84	& 211694.22 \\ 
        PING\_MBULK	  & 157978.9 	& 220975.64	& 218040.66 \\ 
        SET	          & 153390.59 	& 211647.77	& 210302.41 \\ 
        GET	          & 155994.34 	& 218670.04	& 219300.26 \\ 
        INCR	      & 152132.93 	& 219217.17	& 219302.27 \\ 
        LPUSH	      & 149886.75 	& 211691.68	& 211960.12 \\ 
        RPUSH	      & 150504.81 	& 214605.21	& 214053.79 \\ 
        LPOP	      & 148347.63 	& 209365.11	& 209308.7 \\ 
        RPOP	      & 150713.57 	& 210426.48	& 210138.93 \\ 
        SADD	      & 156514.14 	& 217682.08	& 217877.7 \\ 
        HSET	      & 152276.36 	& 209336.32	& 211663.65 \\ 
        SPOP	      & 157350.85 	& 217016.38	& 221988.13 \\ 
        ZADD	      & 149385.73 	& 206069.24	& 207479.97 \\ 
        ZPOPMIN	      & 158360.84 	& 219783.59	& 221895.17 \\ 
        LRANGE\_100   & 92696.06	& 114471.67 & 113062.24\\ 
        LRANGE\_300   & 39268.41	& 39732.19  & 39629.18\\ 
        LRANGE\_500   & 27429.67	& 27843.41  & 27338.37\\ 
        LRANGE\_600   & 23876.49	& 23649.05  & 23674.88\\ 
        MSET (10 keys)& 125746.68 	& 160040.56	& 157920.17 \\ 
        \hline
    \end{tabular}
    \end{threeparttable}
    \caption{Complete results of redis-benchmark.}
    \label{app:complete:redis}
\end{table}

\section{Complete results for SQLite benchmark}
\label{appendix:sqlite-full-results}

Table~\ref{app:complete:sqlite} displays the complete results from testing SQLite with speedtest1. The first two columns indicate the test numbers along with their corresponding test names. The last three columns show the performance metrics for Linux, \aster, and \aster without IOMMU, measured in seconds.

\begin{table}[H]
    \centering
    \scriptsize
    \begin{tabular}{p{0.43cm}|p{3.4cm}|p{0.65cm}<{\centering}p{0.68cm}<{\centering}c}
    \hline
    \textbf{Num.}
         & \textbf{Test Name} & 
         \makecell{\textbf{Linux} \\ (s)} & 
         \makecell{\textbf{Aster.} \\ (s)} & 
         \makecell{\textbf{Aster. no}\\ \textbf{IOMMU(s)}}\\
        \hline
        100 & 500000 INSERTs into table with no index &	         0.27 	& 0.33 	& 0.32 \\ 
        110 & 500000 ordered INSERTS with one index/PK &	     0.43 	& 0.49 	& 0.49 \\ 
        120 & 500000 unordered INSERTS with one index/PK &	     0.88 	& 1.00 	& 1.00 \\ 
        130 & 25 SELECTS, numeric BETWEEN, unindexed &	         0.40 	& 0.45 	& 0.44 \\ 
        140 & 10 SELECTS, LIKE, unindexed &	                     0.61 	& 0.71 	& 0.73 \\ 
        142 & 10 SELECTS w/ORDER BY, unindexed &	             1.17 	& 1.35 	& 1.34 \\ 
        145 & 10 SELECTS w/ORDER BY and LIMIT, unindexed &	     0.49 	& 0.57 	& 0.56 \\ 
        150 & CREATE INDEX five times &	                         0.95 	& 1.16 	& 1.13 \\ 
        160 & 100000 SELECTS, numeric BETWEEN, indexed &	     1.74 	& 2.02 	& 2.03 \\ 
        161 & 100000 SELECTS, numeric BETWEEN, PK &	             1.75 	& 2.02 	& 2.02 \\ 
        170 & 100000 SELECTS, text BETWEEN, indexed &	         1.72 	& 2.06 	& 2.03 \\ 
        180 & 500000 INSERTS with three indexes &	             2.14 	& 2.41 	& 2.42 \\ 
        190 & DELETE and REFILL one table &                      2.09 	& 2.38 	& 2.38 \\ 
        200 & VACUUM &                                           1.59 	& 2.21 	& 2.07 \\ 
        210 & ALTER TABLE ADD COLUMN, and query &	             0.04 	& 0.04 	& 0.04 \\ 
        230 & 100000 UPDATES, numeric BETWEEN, indexed &	     1.81 	& 2.11 	& 2.08 \\ 
        240 & 500000 UPDATES of individual rows &	             1.34 	& 1.58 	& 1.55 \\ 
        250 & One big UPDATE of the whole 500000-row table &	 0.21 	& 0.26 	& 0.24 \\ 
        260 & Query added column after filling &	             0.02 	& 0.02 	& 0.02 \\ 
        270 & 100000 DELETEs, numeric BETWEEN, indexed &	     2.26 	& 2.63 	& 2.58 \\ 
        280 & 500000 DELETEs of individual rows &	             2.19 	& 2.6 	& 2.58 \\ 
        290 & Refill two 500000-row tables using REPLACE &	     3.85 	& 4.31 	& 4.22 \\ 
        300 & Refill a 500000-row table using (b\&1)==(a\&1) &	 2.20 	& 2.51 	& 2.48 \\ 
        310 & 100000 four-ways joins &	                         3.60 	& 4.27 	& 4.25 \\ 
        320 & subquery in result set &	                         7.14 	& 8.3 	& 8.35 \\ 
        400 & 700000 REPLACE ops on an IPK &	                 1.44 	& 1.57 	& 1.58 \\ 
        410 & 700000 SELECTS on an IPK &	                     2.25 	& 3.06 	& 3.05 \\ 
        500 & 700000 REPLACE on TEXT PK &	                     1.66 	& 1.82 	& 1.85 \\ 
        510 & 700000 SELECTS on a TEXT PK &	                     2.56 	& 3.4 	& 3.41 \\ 
        520 & 700000 SELECT DISTINCT &	                         0.57 	& 0.62 	& 0.64 \\ 
        980 & PRAGMA integrity\_check &	                         3.33 	& 3.95 	& 3.97 \\ 
        990 & ANALYZE &	                                         0.20 	& 0.22 	& 0.22 \\ 
        \hline
            & TOTAL &	                                         52.88 	& 62.44 	& 62.07 \\ 
        \hline
    \end{tabular}
    \caption{Complete results of SQLite.}
    \label{app:complete:sqlite}
\end{table}
